\documentclass[twocolumn]{autart}

\usepackage{amsmath,bm}
\usepackage{amsfonts}
\usepackage{amssymb}
\usepackage{graphicx}
\usepackage{color}
\usepackage{caption}
\usepackage{subcaption}
\usepackage{cite}
\usepackage{multicol}
\usepackage[T1]{fontenc}

\usepackage[mathlines, switch]{lineno}

\usepackage{etoolbox} 

\newcommand*\linenomathpatchAMS[1]{%
  \expandafter\pretocmd\csname #1\endcsname {\linenomathAMS}{}{}%
  \expandafter\pretocmd\csname #1*\endcsname{\linenomathAMS}{}{}%
  \expandafter\apptocmd\csname end#1\endcsname {\endlinenomath}{}{}%
  \expandafter\apptocmd\csname end#1*\endcsname{\endlinenomath}{}{}%
}

\expandafter\ifx\linenomath\linenomathWithnumbers
  \let\linenomathAMS\linenomathWithnumbers
  \patchcmd\linenomathAMS{\advance\postdisplaypenalty\linenopenalty}{}{}{}
\else
  \let\linenomathAMS\linenomathNonumbers
\fi

\linenomathpatchAMS{gather}
\linenomathpatchAMS{multline}
\linenomathpatchAMS{align}
\linenomathpatchAMS{alignat}
\linenomathpatchAMS{flalign}

\usepackage{textcomp}
\usepackage{times} 
\usepackage{hyperref}
\usepackage{algorithm} 
\usepackage{savesym}
\savesymbol{AND}
\savesymbol{OR}
\savesymbol{NOT}
\savesymbol{TO}
\savesymbol{COMMENT}
\savesymbol{BODY}
\savesymbol{IF}
\savesymbol{ELSE}
\savesymbol{ELSIF}
\savesymbol{FOR}
\savesymbol{WHILE}
\usepackage{algorithmic}

\usepackage{cases}

\usepackage[mathscr]{euscript}
\usepackage{mathrsfs}

\usepackage[numbers,sort&compress]{natbib}

\usepackage{tikz}
\usetikzlibrary{positioning}
\usetikzlibrary{calc}
\usetikzlibrary{shapes}
\usetikzlibrary{arrows.meta}
\usetikzlibrary{decorations.text}
\usetikzlibrary{matrix}

\definecolor{myRed}{RGB}{192,0,0}
\definecolor{myBlue}{RGB}{0,112,192}

\def\centerarc[#1](#2)(#3:#4:#5)
{ \draw[#1] ($(#2)+({#5*cos(#3)},{#5*sin(#3)})$) arc (#3:#4:#5); }

\newtheorem{definition}{Definition}
\newtheorem{problem}{Problem}
\newtheorem{assumption}{Assumption}
\newtheorem{example}{Example}
\newtheorem{theorem}{Theorem}
\newtheorem{lemma}{Lemma}

\newtheorem{corollary}{Corollary}

\usepackage[labelformat=simple]{subcaption}

\begin{document}

\begin{frontmatter}
\title{Edge Localization in Two Dimensional Space via Orientation Estimation}

\author[gist]{Koog-Hwan Oh},~
\author[uow]{Bar{\i}{\c{s}} Fidan},~
\author[gist]{Hyo-Sung Ahn\corauthref{cor}}
\corauth[cor]{Corresponding author. \emph{Email address:} \textit{hyosung@gist.ac.kr}}
\address[gist]{School of Mechanical Engineering, Gwangju Institute of Science and Technology (GIST), Gwangju, Republic of Korea}
\address[uow]{Department of Mechanical and Mechatronics Engineering, University of Waterloo, Waterloo, ON, N2L 3G1, Canada}

\begin{keyword}
Localization, Multi-agent systems, Orientation estimation, Network systems, Bearing vectors, Subtended angles
\end{keyword}

\begin{abstract}
This paper focuses on the problem of estimating bearing vectors between the agents in a two dimensional multi-agent network based on subtended angle measurements, called \textit{edge localization} problem.
We propose an edge localization graph to investigate the solvability of this problem and a distributed estimation method via orientation estimation of virtual agents to solve the problem.
Under the proposed method, the estimated bearing vector exponentially converges to the real one with a common bias if and only if the edge localization graph has an oriented spanning tree.
Furthermore, the estimated variables exponentially converge to the true values if the edge localization graph has an oriented spanning tree with a root knowing the bearing vector from it to one of its neighbors.

\end{abstract}

\end{frontmatter}
\section{Introduction}\label{sec:Introduction}
Recently, multi-agent systems have attracted attention of researchers because they have potential applications in various areas.
In particular, theoretical challenges have remarkably emerged in the formation control problems \cite{anderson2011maintaining,oh2011formation,sun2016exponential,ren2005consensus}.
Network localization in multi-agent systems has also been in the limelight of scientific interest owing to its usefulness in various sensor network and formation control applications \cite{aspnes2006theory,ahn2009Formation}.

The formation control of multi-agent systems was approached via several schemes, which are categorized according to types of sensed and controlled variables \cite{oh2015survey}.
Among them, angle-based formation control schemes use angle-based measurements, such as bearing and subtended angle measurements between the agents, or aims to achieve desired formation shape defined by angle-based variables.
In the work of Trinh et al. \cite{trinh2018formations}, the angle-based formation control for a group of single-integrator agents with directed cycle sensing topology has been studied.
They have showed that the desired bearing-based formation could be achieved using only bearing measurements in both two and three dimensional space.
The works in \cite{basiri2010distributed,bishop2011very} have studied formation controls of three or four agents in two dimensional space based only on the bearing measurements and subtended angle constraints.
Zhao et al. \cite{zhao2014distributed} have further studied formation control of arbitrary number of agents with an undirected cycle interaction.
The results in \cite{basiri2010distributed,bishop2011very,zhao2014distributed} also showed that the multi-agent systems can be controlled to achieve the desired formations defined by a set of desired subtended angles using only bearing measurements.
Formation control with multiple subtended angle constraints and a single distance constraint has been introduced in \cite{bishop2012control}.
The formation control task in \cite{bishop2012control} could be achieved using a mixture of bearing and range measurements.

Network localization problems generally target estimating the positions of agents.
In common with the formation control, they can be conducted via angle-based measurements.
Rong and Sichitiu \cite{rong2006angle} proposed a network localization scheme based on angle-of-arrival (AOA) measurements.
They showed that their proposed method achieves network localization using relative bearing measurements.
The work in \cite{bishop2009Bearing} studied optimal localization of a single agent, in both two and three dimensional space, based on the bearing measurements from the other agents.
A localizability condition for the bearing-based network localization in two dimensional space was proposed in \cite{zhu2014adistributed}.
Recently, Zhao and Zelazo \cite{zhao2016localizability} proposed an extension of this condition to arbitrary dimensional spaces.
In network localization, estimation of the orientations of agents can be also considered.
The work in \cite{piovan2013onframe} addressed an orientation localization problem, to compute the orientation of each node based on the relative bearing measurements, and developed an algorithm for solving this problem.

This paper defines and analyzes a new localizaton problem, the problem of estimating the bearing vectors between the agents in a two dimensional multi-agent network with only the subtended angle measurements.
We call this problem as \textit{edge localization problem}.
The existing angle-based formation control and network localization schemes mentioned above depend on the bearing measurements,
because agents in formation control can not be controlled or states of agents in network localization can not be estimated without the bearing measurements.
The bearing information is even required to control the agents in the formation control tasks defined with only subtended angle constraints as in \cite{basiri2010distributed,bishop2011very,zhao2014distributed}.
In certain practical cases, it might be difficult to measure bearing vectors.
For example, in vision-based formation control \cite{moshtagh2009vision} or in landmark-based navigation \cite{hayet2007visual}, it is more natural to measure subtended angles between two neighboring agents.
In such cases, the subtended angles should be transformed into bearing vectors in order to achieve a bearing-based formation control.  
Then, via such a transformation, the angle-based formation control or network localization without the bearing measurements could be conducted.
Moreover, such transformation can be used to enhance the bearing measurements with the subtended angle measurements if the subtended angles and the bearing vectors are independently measured.
Consequently, as the main contributions of this paper, first we define the \textit{edge localization problem} in two dimensional space and propose an edge localization graph to investigate its solvability.
Next, we propose virtual agents and their interaction graph, called \textit{localization interaction graph}, to solve the problem.
Finally, we propose and analyze a distributed method for achieving the edge localization.

The remainder of this paper is organized in the following manner.
In Section 2, some preliminaries of graph theory and consensus properties are presented and the edge localization problem is formulated.
Section 3 introduces the edge localization graph and the localization interaction graph.
In Section 4, our method for edge localization via orientation estimation is proposed.
In Section 5, some simulations are performed to validate our proposed method.
Finally, Section 6 concludes this paper and outlines directions for future works.


\section{Preliminaries and Problem Statement}\label{sec:preliminaries}
\subsection{Preliminaries}\label{subsec:preliminaries}
A directed graph $\mathcal{G} = \left( \mathcal{V}, \mathcal{E} \right)$ consists of a set of vertices $\mathcal{V}$ and a set of directed edges $\mathcal{E}$.
Each directed edge is an ordered pair $\left(u,v\right)$ of distinct vertices in $\mathcal{V}$.
For a directed edge $\left(u,v\right)$, $v$ is called the head and $u$ is called the tail.
The head and tail of a directed edge $e$ are denoted by $h\left(e\right)$ and $t\left(e\right)$, respectively.
A set of neighbors of $u$ is denoted by $\mathcal{N}_u$, i.e., $v \in \mathcal{N}_u$ if $\left(u,v\right) \in \mathcal{E}$.
For a vertex $u$, $d_{i}\left( u \right)$ and $d_{o}\left( u \right)$ denote the in-degree and out-degree of $u$, respectively.
A directed graph is called symmetric if $\left(u,v\right)$ is a directed edge whenever $\left(v,u\right)$ is.
A directed path is a sequence of ordered edges in a directed graph of the form $\left(v_1,v_2\right), \left(v_2,v_3\right), \ldots, \left(v_{m-1},v_m\right)$.
An oriented spanning tree of a directed graph means a directed tree containing every vertex of the graph \cite{oh2015survey}.
In this paper, the oriented spanning tree is defined as follows:
\begin{definition}[Oriented spanning tree \cite{bidkhori2011bijective}]\label{def:span_t}
Let $\mathcal{G} = \left( \mathcal{V}, \mathcal{E} \right)$ be a directed graph.
An oriented spanning tree of $\mathcal{G}$ is an acyclic subgraph of $\mathcal{G}$ with a root $r$, in which there exists a directed path from every vertex $v \in \mathcal{V}\setminus \left\{ r \right\}$ to the root $r$.
\end{definition}
A directed line graph is defined in Definition \ref{def:line_g} and Figure \ref{fig:line_exp1} shows an example of a directed graph and its directed line graph.
\begin{definition}[Directed line graph \cite{bidkhori2011bijective}] \label{def:line_g}
Let $\mathcal{G} = \left( \mathcal{V}, \mathcal{E} \right)$ be a directed graph.
The directed line graph of $\mathcal{G}$ is a directed graph with vertex set $\mathcal{E}$, and with a directed edge $\left( e_1, e_2 \right)$ for every pair of edges $e_1$ and $e_2$ of $\mathcal{G}$ with $h\left( e_1 \right) = t\left( e_2 \right)$.
We denote the directed line graph of $\mathcal{G} = \left( \mathcal{V}, \mathcal{E} \right)$ by $L\left(\mathcal{G}\right) = \left( \mathcal{E}, L\left( \mathcal{E}\right) \right)$ where $L\left( \mathcal{E}\right) = \left\{ \left( e_1, e_2 \right) \mid h\left( e_1 \right) = t\left( e_2 \right), \forall e_1, e_2 \in \mathcal{E} \right\}$.  
\end{definition}
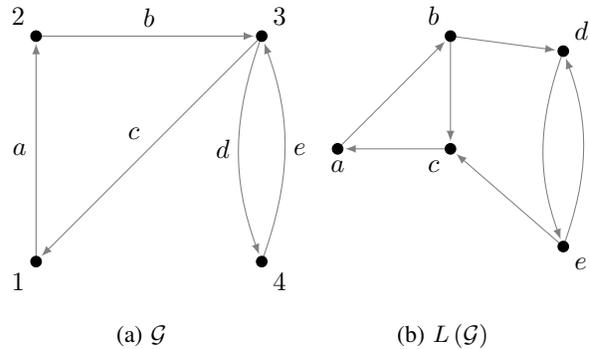
\begin{figure}[!h]
	\centering
	\begin{subfigure}{0.45\columnwidth}
		\centering

		\begin{tikzpicture}
		\useasboundingbox (0,0) rectangle (4,4);

		\coordinate (v1) at (0.5,0.5);
		\coordinate (v2) at (0.5,3.5);
		\coordinate (v3) at (3.5,3.5);
		\coordinate (v4) at (3.5,0.5);
		
		\draw[gray,-latex,shorten >= 0.1cm] (v1) -- (v2)
		node[pos=0.5,left,black]{$a$};
		\draw[gray,-latex,shorten >= 0.1cm] (v2) -- (v3)
		node[pos=0.5,above,black]{$b$};
		\draw[gray,-latex,shorten >= 0.1cm] (v3) -- (v1)
		node[pos=0.5,above left,black]{$c$};
		
		\draw[gray, latex-, shorten <= 0.1cm] (v4) to[out=110,in=-110] (v3);
		\node[black,left] (d) at (3.2,2) {$d$};
		
		\draw[gray, latex-, shorten <= 0.1cm] (v3) to[out=-70,in=70] (v4);
		\node[black,right] (e) at (3.8,2) {$e$};

		\draw [fill=black] (v1) circle (2pt);
		\draw [fill=black] (v2) circle (2pt);
		\draw [fill=black] (v3) circle (2pt);
		\draw [fill=black] (v4) circle (2pt);

		\node[below left = 1pt of v1] (text_v1) {$1$};
		\node[above left = 1pt of v2] (text_v2) {$2$};
		\node[above right = 1pt of v3] (text_v3) {$3$};
		\node[below right = 1pt of v4] (text_v4) {$4$};
		
		\end{tikzpicture}
		
		\caption{$\mathcal{G}$} \label{fig:line_exp1_a}
	\end{subfigure}~
	\begin{subfigure}{0.45\columnwidth}
		\centering

		\begin{tikzpicture}
		\useasboundingbox (0,0) rectangle (4,4);
		
		\coordinate (a) at (0.5,2);
		\coordinate (b) at (2,3.5);
		\coordinate (c) at (2,2);
		\coordinate (d) at (3.5,3.3);
		\coordinate (e) at (3.5,0.7);
		
		\draw[gray,-latex,shorten >= 0.1cm] (a) -- (b);
		\draw[gray,-latex,shorten >= 0.1cm] (b) -- (c);
		\draw[gray,-latex,shorten >= 0.1cm] (c) -- (a);
		\draw[gray,-latex,shorten >= 0.1cm] (b) -- (d);
		\draw[gray,-latex,shorten >= 0.1cm] (e) -- (c);
		
		\draw[gray, latex-, shorten <= 0.1cm] (e) to[out=110,in=-110] (d);
		\draw[gray, latex-, shorten <= 0.1cm] (d) to[out=-70,in=70] (e);

		\draw [fill=black] (a) circle (2pt);
		\draw [fill=black] (b) circle (2pt);
		\draw [fill=black] (c) circle (2pt);
		\draw [fill=black] (d) circle (2pt);
		\draw [fill=black] (e) circle (2pt);
		
		\node[black,below = 1pt of a] (text_a) {$a$};
		\node[black,above left = 1pt of b] (text_b) {$b$};
		\node[black,below left = 1pt of c] (text_c) {$c$};
		\node[black,above right = 1pt of d] (text_d) {$d$};
		\node[black,below right = 1pt of e] (text_e) {$e$};
		
		\end{tikzpicture}

		\caption{$L\left(\mathcal{G}\right)$} \label{fig:line_exp1_b}
	\end{subfigure}
	\caption{An example of a directed graph $\mathcal{G}$ and its directed line graph $L\left(\mathcal{G}\right)$.}
	\label{fig:line_exp1}
\end{figure}
To prevent ambiguities between edges of $\mathcal{G}$ and the equivalent vertices of $L\left( \mathcal{G} \right)$, in the sequel, we denote each $\left( u,v \right) \in \mathcal{E}$ by $\left( u,v \right)_e$, called edge node, if it refers to a vertex of $L\left( \mathcal{G} \right)$.

Consider a network of $N$ agents whose interaction graph is a directed weighted graph $\mathcal{G} = \left( \mathcal{V}, \mathcal{E}, \mathcal{A} \right)$ where $\mathcal{A}$ is a weighted adjacency matrix with nonnegative elements denoted by $a_{ij}$ which is assigned to $\bigl( i, j\bigr) \in \mathcal{E}$.
The Laplacian matrix of the directed weighted graph $\mathcal{G}$ is defined as $\mathcal{L} = \left[{ l_{ij} }\right]$, where 
\begin{align*}
{ l_{ij} = } \left.
	\begin{cases}
	{\sum\limits_{k \in \mathcal{N}_{i}} {a_{ik}}}	& {\text{if } i = j}, \\
	{-a_{ij}}					& {\text{if } j \in \mathcal{N}_{i}},	\\
	{0}						& {\text{otherwise}}.
	\end{cases}
\right.
\end{align*}
The following lemma is well-known \cite{ren2005consensus}.
\begin{lemma}[\cite{ren2005consensus}]\label{lem:eigen}
The directed weighted graph $\mathcal{G}$ has an oriented spanning tree if and only if the Laplacian matrix $\mathcal{L}$ associated to $\mathcal{G}$ has a simple zero eigenvalue with right eigenvector $\bold{1}_{N} = \left[ 1, \ldots, 1\right]^{T}$.
Then, all the other eigenvalues have strictly positive real parts.
\end{lemma}
A classical consensus protocol in continuous-time can be simply represented as follows \cite{scardovi2009brief}: 
\begin{align}\label{eq:cl_con}
\dot{\bold{x}}_{i}\left(t\right) = \sum\limits_{j \in \mathcal{N}_{i}}{a_{ij} \left( \bold{x}_{j}\left(t\right) - \bold{x}_{i}\left(t\right) \right)}, \quad \forall i \in \mathcal{V}
\end{align}
where $\bold{x}_{i} \in \mathbb{R}^n$ is a column vector.
The equation (\ref{eq:cl_con}) can be rewritten using the Laplacian matrix as
\begin{align}\label{eq:consensus1}
\dot{\bold{x}}\left(t\right) = - \left( \mathcal{L} \otimes I_n \right) \bold{x}\left(t\right)
\end{align}
where $\bold{x} = \left[{ {\bold{x}}_{1}^T, {\bold{x}}_{2}^T,  \ldots, {\bold{x}}_{N}^T}\right]^T$ and $A \otimes B$ denotes the Kronecker product of the matrices.
The following lemma states that a necessary and sufficient condition for exponential stability of (\ref{eq:consensus1}) is $\mathcal{G}$ having an oriented spanning tree \cite{lee2016distributed}.
\begin{lemma}[\cite{lee2016distributed}] \label{lem:consensus}
The equilibrium set $\mathscr{E}_{\mathcal{X}} = \bigl\{ \bold{x} \in \mathbb{R}^{nN} : \bold{x}_{i} = \bold{x}_{j}, \forall i,j \in \mathcal{V} \bigr\}$ of (\ref{eq:consensus1}) is exponentially stable if and only if $\mathcal{G}$ has an oriented spanning tree.
Furthermore, the state $\bold{x}\left(t\right)$ converges to a finite point in $\mathscr{E}_{\mathcal{X}}$.
\end{lemma}


\subsection{Problem Statement} \label{subsec:problem}

\begin{figure}[h]
	\centering
	\begin{subfigure}[t]{0.48\columnwidth}
		\centering

		\begin{tikzpicture}
		\useasboundingbox (-0.75,-0.75) rectangle (3.75,3.75);
	    \draw [<->,very thin] (-0.5,3.5) node (yaxis) {}
        |- (3.5,-0.5) node (xaxis) [below left] {${}^{g}\sum$};		
		
		\coordinate (u) at (0.5, 0.5);
		\coordinate (v) at (3,1);
		\coordinate (w) at (1,3);
		
		\draw[gray,thin,dashed] (u) -- (v);
		\draw[gray,thin,dashed] (u) -- (w);

		\centerarc[black,thin,->](u)(13:77:0.5)

		\node (gvu) at ($(u)!0.35!(v)$) {};
		\node (gwu) at ($(u)!0.35!(w)$) {};
		
		\draw[black,-latex] (u) -- (gvu);
		\draw[black,-latex] (u) -- (gwu);
		
		\node (guv) at ($(u)!0.65!(v)$) {};
		\node (guw) at ($(u)!0.65!(w)$) {};
		
		\draw[black,-latex] (v) -- (guv);
		\draw[black,-latex] (w) -- (guw);

		\draw [fill=black] (u) circle (2pt);
		\draw [fill=black] (v) circle (2pt);
		\draw [fill=black] (w) circle (2pt);

		\node[below left = 1pt of u] (text_u) {$u$};
		\node[below right= 1pt of v] (text_v) {$v$};
		\node[above right= 1pt of w] (text_w) {$w$};
		
		\node (text_awv) at (1.7,1.15) {$\alpha_{wv}^{u}$ = - $\alpha_{vw}^{u}$};

		\node[black, below = 0.1pt of gvu] (text_gvu) {$\bold{g}_{vu}$};
		\node[black, below right = 0.1pt of guv] (text_gvu) {$\bold{g}_{uv}$};
		\node[black, left = 0.1pt of gwu] (text_gwu) {$\bold{g}_{wu}$};
		\node[black, above right = 0.1pt of guw] (text_guw) {$\bold{g}_{uw}$};

		\end{tikzpicture}
		
		\caption{A pair of subtended angles and bearing vectors.}	\label{fig_sub_and_bearings}
	\end{subfigure}~
	\begin{subfigure}[t]{0.48\columnwidth}
		\centering

		\begin{tikzpicture}
		\useasboundingbox (-0.75,-0.75) rectangle (3.75,3.75);

		\coordinate (u) at (0.5, 0.5);
		\coordinate (v) at (3,1);
		\coordinate (w) at (1,3);

		\draw [gray, -latex, shorten >= 0.1cm] (u) to[out=25,in=170] (v);
		\draw [gray, -latex, shorten >= 0.1cm] (v) to[out=-150,in=-5] (u);

		\draw [gray, -latex, shorten >= 0.1cm] (u) to[out=60,in=-80] (w);
		\draw [gray, -latex, shorten >= 0.1cm] (w) to[out=-120,in=95] (u);
		
		\draw [fill=black] (u) circle (2pt);
		\draw [fill=black] (v) circle (2pt);
		\draw [fill=black] (w) circle (2pt);

		\node[below left = 1pt of u] (text_u) {$u$};
		\node[below right= 1pt of v] (text_v) {$v$};
		\node[above right = 1pt of w] (text_w) {$w$};
		
		\end{tikzpicture}

		\caption{An edge set $\mathcal{E}$ of (a)}
	\end{subfigure}
	\caption{Description of (a) A pair of subtended angles and bearing vectors related to them and (b) An edge set $\mathcal{E}$ determined by a pair of the subtended angle and Assumption \ref{assum:1}.}
	\label{fig:assum1}
\end{figure}
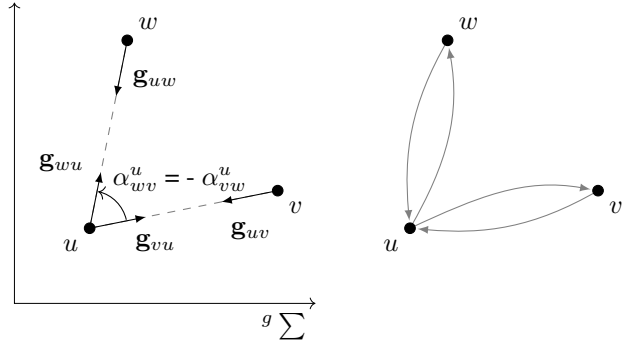

Consider a network of $N$ agents in two dimensional space and a given set of subtended angle measurements denoted by $\mathcal{S}$.
Let $\mathcal{V} = \left\{ 1, \ldots, N \right\}$ and $\mathcal{E}$ denote the set of agents and directed edges, respectively.
The subtended angle from agent $v$ to agent $w$ measured at agent $u$ is defined by $\alpha_{wv}^{u} \in \left[ -\pi, \pi \right)$, in counter-clock wise direction.
Note that if an agent $u$ measures a subtended angle $\alpha_{wv}^{u}$, then it also has the measurement of $\alpha_{vw}^{u} = - \alpha_{wv}^{u}$, i.e., $\alpha_{vw}^{u} \in \mathcal{S}$ if $\alpha_{wv}^{u} \in \mathcal{S}$.
The bearing vector from agent $u$ to agent $v$ is defined as
\begin{align*}
{\bold{g}_{vu}} \triangleq \frac{{{\bold{p}_v} - {\bold{p}_u}}}{{\left\| {{\bold{p}_v} - {\bold{p}_u}} \right\|}},
\end{align*}
where $\bold{p}_u\in\mathbb{R}^2$ denotes the position of agent $u$ with respect to the global coordinate frame denoted by ${}^{g}\sum$.
The subtended angle $\alpha_{wv}^{u}$ can be considered as the angle from the directed edge $\left( u,v \right)$ to the directed edge $\left( u,w \right)$.
In other words, in addition to the subtended angle $\alpha_{wv}^{u}$, if $\angle{\bold{g}_{wu}}$ or $\angle{\bold{g}_{uw}}$ is known,
the angles $\angle{\bold{g}_{uv}}$ and $\angle{\bold{g}_{vu}}$ can be calculated.
But, they can not be calculated with just $\alpha_{wv}^{u}$ known.
A pair of subtended angles and the corresponding bearing vectors are illustrated in Figure \ref{fig_sub_and_bearings}.
Due to the relationship between the subtended angles and the bearing vectors, we assume the directed edge set $\mathcal{E}$ as follows:
\begin{assumption}\label{assum:1}
If the measurements of the subtended angles $\alpha_{wv}^{u}$ and $\alpha_{vw}^{u}$ are available, i.e., if $\alpha_{wv}^u \in \mathcal{S}$ (and hence  $\alpha_{vw}^u \in \mathcal{S}$),
then $\left(u,v\right)$, $\left(v,u\right)$, $\left(u,w\right)$, $\left(w,u\right) \in \mathcal{E}$.
\end{assumption}

The problem of interest in this paper is to estimate the bearing vectors between the agents based on the given set of subtended angle measurements, $\mathcal{S}$.
It is typically desired to conduct the edge localization in a distributed way.
As illustrated in Figure \ref{fig_sub_and_bearings}, the subtended angles $\alpha_{wv}^{u}$ and $\alpha_{vw}^{u}$ are geometrically related to the bearing vectors $\bold{g}_{vu}$, $\bold{g}_{wu}$, $\bold{g}_{uv}$, and $\bold{g}_{uw}$.
The distributed approaches based only on subtended angle measurements require communication among agents.
Hence, we consider the interaction topology among agents as the communication topology.
The communications between agents need to be bi-directional.
Thus, even though the basic underlying graph is directed one in Definition \ref{def:span_t}, the communication graph is assumed to be bi-directional in a directed setup.
So, without a notational confusion, we suppose that the communication graph is a connected graph.
The edge localization problem is then stated as follows:
\begin{problem}[Edge localization]\label{prob:EL}
Consider $N$ agents, whose communication graph is $\mathcal{G} = \left( \mathcal{V}, \mathcal{E} \right)$, with a set $\mathcal{S}$ of subtended angle measurements under Assumption \ref{assum:1}.
Let $\mathcal{G}$ be a connected graph. Find the bearing vector $\bold{g}_{vu}$, for each edge $\left(u,v\right) \in \mathcal{E}$, using communications in $\mathcal{G}$ and the measurement set $\mathcal{S}$.
\end{problem}


\section{Edge Localization Graph and Localization Interaction Graph} \label{sect:IG}

Consider the setting of Problem \ref{prob:EL} with a network of $N$ agents and their connected communication graph $\mathcal{G} = \left( \mathcal{V}, \mathcal{E} \right)$.
Note that the communication graph $\mathcal{G}$ is symmetric since it is determined by the set $\mathcal{S}$ of subtended angle measurements and Assumption \ref{assum:1}.
Let $\angle{\bold{g}_{vu}}\in \left[ -\pi, \pi \right)$ be an angle of the bearing vector $\bold{g}_{vu}$ with respect to the global coordinate frame ${}^{g}\sum$.
Then, for each edge $\left(u,v\right) \in \mathcal{E}$, estimating the angle $\angle{\bold{g}_{vu}}$ is identical to finding the bearing vector $\bold{g}_{vu}$ in Problem \ref{prob:EL},
since 
\begin{align}
\bold{g}_{vu} = \left[  \cos \left(\angle{\bold{g}_{vu}}\right), \sin \left(\angle{\bold{g}_{vu}}\right)  \right]^T.
\end{align}
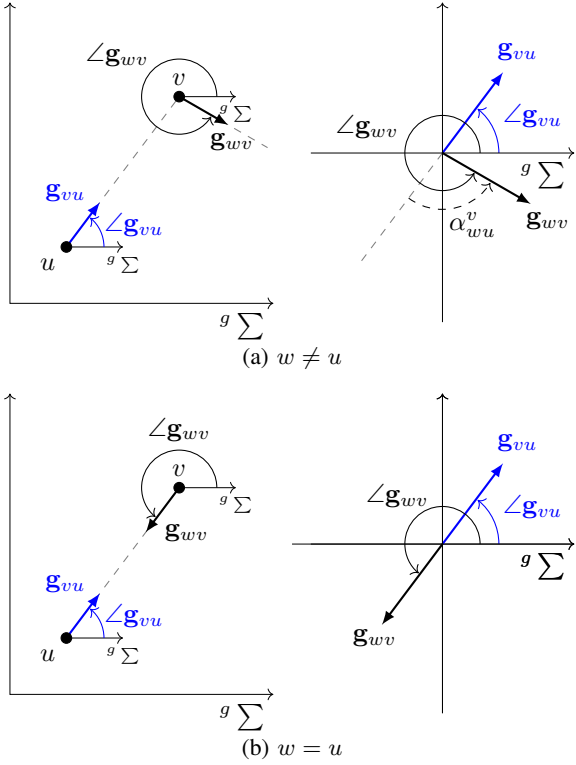
\begin{figure}[!h]
	\centering
	\begin{subfigure}{1\columnwidth}
		\centering
		\begin{tikzpicture}
		\useasboundingbox (0,0) rectangle (8,4.5);

	    \draw[<->,very thin] (0.25,4.25) node (yaxis1) {}
        |- (3.75,0.25) node (xaxis1) [below left] {${}^{g}\sum$};

		\coordinate (u) at (1, 1);
		\coordinate (v) at (2.5, 3);

		\node at (120:1cm) (gvu) {};
		\node (gvu) at ([shift=({53.13:0.9})]u) {};
		\node (gwv) at ([shift=({-30:0.9})]v) {};
		\node (w) at ([shift=({-30:1.5})]v) {};

		\draw[gray,thin,dashed] (u) -- (v);
		\draw[gray,thin,dashed] (v) -- (w);

  	    \draw[-latex,blue,thick] (u) -- (gvu);
  	    \draw[-latex,black,thick] (v) -- (gwv);

  	    \draw[->,very thin] (u) -- +(0.75,0) node (xaxisu) [font=\tiny,below] {${}^{g}\sum$};		
  	    \draw[->,very thin] (v) -- +(0.75,0) node (xaxisv) [font=\tiny,below] {${}^{g}\sum$};		
		
		\centerarc[blue,thin,->](u)(0:51:0.5)
		\centerarc[black,thin,->](v)(0:328:0.5)

		\draw [fill=black] (u) circle (2pt);
		\draw [fill=black] (v) circle (2pt);
		
		\node[below left = 1pt of u] (text_u) {$u$};		
		\node[above = 1pt of v] (text_v) {$v$};

		\node[blue,left = 1pt of gvu] (text_gvu) {$\bold{g}_{vu}$};
		\node[black] (text_gwv) at (3.2,2.4) {$\bold{g}_{wv}$};

		\node[blue] (text_agvu) at (1.9,1.3) {$\angle{\bold{g}_{vu}}$};
		\node[black,above left = 0.4cm of v] (text_agwv) {$\angle{\bold{g}_{wv}}$};

		\node (origin) at (6,2.25) {};

		\node (guv) at ([shift=({-126.87:2})]origin) {};
		\node (gvu) at ([shift=({53.13:1.5})]origin) {};
		\node (gwv) at ([shift=({-30:1.5})]origin) {};
		
		\draw[gray,thin,dashed] (6,2.25) -- (guv);
  	    \draw[-latex,blue,thick] (6,2.25) -- (gvu);
   	    \draw[-latex,black,thick] (6,2.25) -- (gwv);
  	   
   		\centerarc[blue,thin,->](origin)(0:51.5:0.75)
   		\centerarc[black,thin,->](origin)(0:328:0.5)
   		\centerarc[black,thin,->,dashed](origin)(233.13:328:0.75)

  	    \draw[->,very thin] (6,0) -- (6,4.25) node (yaxis2) {};
		\draw[->,very thin] (4.25,2.25) -- (7.75,2.25) node (xaxis2) [below left] {${}^{g}\sum$};  	    
  	    
   		\node[blue] at (7,3.6) {$\bold{g}_{vu}$};
   		\node[blue] at (7.2,2.75) {$\angle{\bold{g}_{vu}}$};
   		\node[black] at (7.4,1.35) {$\bold{g}_{wv}$};
   		\node[black] at (5,2.6) {$\angle{\bold{g}_{wv}}$};
   		\node[black] at (6.4,1.3) {$\alpha_{wu}^{v}$};

		\end{tikzpicture}
		\caption{$w \ne u$} \label{fig:geo_re_a}
	\end{subfigure}\\
	\begin{subfigure}{1\columnwidth}
		\centering
		\begin{tikzpicture}
		\useasboundingbox (0,0) rectangle (8,4.5);

	    \draw[<->,very thin] (0.25,4.25) node (yaxis1) {}
        |- (3.75,0.25) node (xaxis1) [below left] {${}^{g}\sum$};

		\coordinate (u) at (1, 1);
		\coordinate (v) at (2.5, 3);

		\node at (120:1cm) (gvu) {};
		\node (gvu) at ([shift=({53.13:0.9})]u) {};
		\node (gwv) at ([shift=({-126.87:0.9})]v) {};

		\draw[gray,thin,dashed] (u) -- (v);

  	    \draw[-latex,blue,thick] (u) -- (gvu);
  	    \draw[-latex,black,thick] (v) -- (gwv);

  	    \draw[->,very thin] (u) -- +(0.75,0) node (xaxisu) [font=\tiny,below] {${}^{g}\sum$};		
  	    \draw[->,very thin] (v) -- +(0.75,0) node (xaxisv) [font=\tiny,below] {${}^{g}\sum$};		
		
		\centerarc[blue,thin,->](u)(0:51:0.5)
		\centerarc[black,thin,->](v)(0:231:0.5)

		\draw [fill=black] (u) circle (2pt);
		\draw [fill=black] (v) circle (2pt);
		
		\node[below left = 1pt of u] (text_u) {$u$};		
		\node[above = 1pt of v] (text_v) {$v$};

		\node[blue,left = 1pt of gvu] (text_gvu) {$\bold{g}_{vu}$};
		\node[black] (text_gwv) at (2.6,2.4) {$\bold{g}_{wv}$};

		\node[blue] (text_agvu) at (1.9,1.3) {$\angle{\bold{g}_{vu}}$};
		\node[black,above = 0.5cm of v] (text_agwv) {$\angle{\bold{g}_{wv}}$};

  	    \draw[->,very thin] (6,0) -- (6,4.25) node (yaxis2) {};
		\draw[->,very thin] (4,2.25) -- (7.75,2.25) node (xaxis2) [below left] {${}^{g}\sum$};

		\node (origin) at (6,2.25) {};

		\node (gvu) at ([shift=({53.13:1.5})]origin) {};
		\node (gwv) at ([shift=({-126.87:1.5})]origin) {};
		
  	    \draw[-latex,blue,thick] (6,2.25) -- (gvu);
   	    \draw[-latex,black,thick] (6,2.25) -- (gwv);
  	   
   		\centerarc[blue,thin,->](origin)(0:51.5:0.75)
   		\centerarc[black,thin,->](origin)(0:231:0.5)

  	    \draw[->,very thin] (6,0) -- (6,4.25) node (yaxis2) {};
		\draw[->,very thin] (4.25,2.25) -- (7.75,2.25) node (xaxis2) [below left] {${}^{g}\sum$};  	    
  	    
   		\node[blue] at (7,3.6) {$\bold{g}_{vu}$};
   		\node[blue] at (7.2,2.75) {$\angle{\bold{g}_{vu}}$};
   		\node[black] at (5.1,1.0) {$\bold{g}_{wv}$};
   		\node[black] at (5.4,2.9) {$\angle{\bold{g}_{wv}}$};

		\end{tikzpicture}
		\caption{$w = u$} \label{fig:geo_re_b}
	\end{subfigure}
	\caption{The geometric relationships between angles of bearing vectors and a subtended angle.}
	\label{fig:geo_re}
\end{figure}

The angle $\angle{\bold{g}_{vu}}$ can be estimated by the geometric relationships between itself and $\angle{\bold{g}_{wv}}$, shown in Figure \ref{fig:geo_re}, as follows:	
\begin{subequations}\label{eq:geo_re}
\begin{numcases}{\angle{\bold{g}_{vu}} = }
{\angle{\bold{g}_{wv}} - \left(\alpha_{wu}^{v} + \pi\right)}	& ${\text{if } w\ne u}$  \nonumber \\ 
{}												& ${\text{ and } \alpha_{wu}^{v} \in \mathcal{S}\text{,}}$ \label{eq:geo_re_1} \\
{\angle{\bold{g}_{wv}} - \pi}					       	& ${\text{if } w   =  u.}$ \label{eq:geo_re_2}
\end{numcases}
\end{subequations}
Our main approach to estimate $\angle{\bold{g}_{vu}}$ for each $\left( u, v \right) \in \mathcal{E}$ is to use the consensus protocol based on the geometric relationships in (\ref{eq:geo_re}), as follows:
First, using the concept of directed line graph in Definition \ref{def:line_g}, each edge $\left( u, v \right) \in \mathcal{E}$ will be mapped to a virtual agent $\left( u,v \right)^{\epsilon}$, called edge agent, whose orientation angle is identical to $\angle{\bold{g}_{vu}}$.
Then each edge agent $\left( u,v \right)^{\epsilon}$ will receive the angle information from its neighbors and estimate its orientation angle through consensus protocol.

For the sake of simplicity and using the geometric relationships between the bearing vectors and the subtended angle measurements, we make the following assumption throughout this paper\footnote{
In a brief version \cite{oh2020distributed} of this paper, there is a mistake omitting this assumption.
So, this full version of \cite{oh2020distributed} and the thesis \cite{oh2020thesis} clearly include it.}:
\begin{assumption}\label{assum:new}
If an agent measures subtended angles between two or more pairs of other agents, all bearing vectors associated with these subtended angle measurements have geometric relationships with each other by them.
\end{assumption}


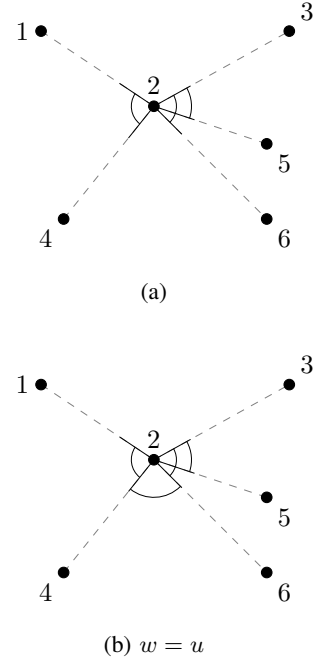
\begin{figure}[!h]
	\centering
	\begin{subfigure}{1\columnwidth}
		\centering
		\begin{tikzpicture}

		\useasboundingbox (0,0) rectangle (4,4);

		\coordinate (v1) at (0.5,3);
		\coordinate (v2) at (2,2);
		\coordinate (v3) at (3.8,3);		
		\coordinate (v4) at (0.8,0.5);
		\coordinate (v5) at (3.5,1.5);
		\coordinate (v6) at (3.5,0.5);

		\draw[gray,very thin,dashed] (0.5,3) -- (2,2);
		\draw[gray,very thin,dashed] (0.8,0.5) -- (2,2);
		\draw[gray,very thin,dashed] (2,2) -- (3.8,3);
		\draw[gray,very thin,dashed] (2,2) -- (3.5,1.5);
		\draw[gray,very thin,dashed] (2,2) -- (3.5,0.5);

		\node (e21) at ([shift=({146.31:0.7})]v2) {};
		\node (e24) at ([shift=({231.34:0.7})]v2) {};
		\node (e23) at ([shift=({29.05:0.7})]v2) {};
		\node (e25) at ([shift=({-18.43:0.7})]v2) {};
		\node (e26) at ([shift=({-45:0.7})]v2) {};		
		\draw[black,very thin] (v2) -- (e21);
		\draw[black,very thin] (v2) -- (e24);
		\draw[black,very thin] (v2) -- (e23);
		\draw[black,very thin] (v2) -- (e25);
		\draw[black,very thin] (v2) -- (e26);

   		\centerarc[black,very thin](v2)(146.31:231.34:0.3)
   		\centerarc[black,very thin](v2)(-18.43:29.05:0.5)
   		\centerarc[black,very thin](v2)(-45:29.05:0.3)
		
		\draw [fill=black] (v1) circle (2pt);
		\draw [fill=black] (v2) circle (2pt);
		\draw [fill=black] (v3) circle (2pt);
		\draw [fill=black] (v4) circle (2pt);
		\draw [fill=black] (v5) circle (2pt);
		\draw [fill=black] (v6) circle (2pt);		

		\node[left = 1pt of v1] (text_v1) {$1$};
		\node[above = 1pt of v2] (text_v2) {$2$};
		\node[above right = 1pt of v3] (text_v3) {$3$};
		\node[below left = 1pt of v4] (text_v4) {$4$};
		\node[below right = 1pt of v5] (text_v5) {$5$};		
		\node[below right = 1pt of v6] (text_v6) {$6$};		
		
		\end{tikzpicture}
		\caption{} \label{fig:EL_example_new_assum_a}
	\end{subfigure}\\
	\begin{subfigure}{1\columnwidth}
		\centering
		\begin{tikzpicture}

		\useasboundingbox (0,0) rectangle (4,4);

		\coordinate (v1) at (0.5,3);
		\coordinate (v2) at (2,2);
		\coordinate (v3) at (3.8,3);		
		\coordinate (v4) at (0.8,0.5);
		\coordinate (v5) at (3.5,1.5);
		\coordinate (v6) at (3.5,0.5);

		\draw[gray,very thin,dashed] (0.5,3) -- (2,2);
		\draw[gray,very thin,dashed] (0.8,0.5) -- (2,2);
		\draw[gray,very thin,dashed] (2,2) -- (3.8,3);
		\draw[gray,very thin,dashed] (2,2) -- (3.5,1.5);
		\draw[gray,very thin,dashed] (2,2) -- (3.5,0.5);

		\node (e21) at ([shift=({146.31:0.7})]v2) {};
		\node (e24) at ([shift=({231.34:0.7})]v2) {};
		\node (e23) at ([shift=({29.05:0.7})]v2) {};
		\node (e25) at ([shift=({-18.43:0.7})]v2) {};
		\node (e26) at ([shift=({-45:0.7})]v2) {};		
		\draw[black,very thin] (v2) -- (e21);
		\draw[black,very thin] (v2) -- (e24);
		\draw[black,very thin] (v2) -- (e23);
		\draw[black,very thin] (v2) -- (e25);
		\draw[black,very thin] (v2) -- (e26);

   		\centerarc[black,very thin](v2)(146.31:231.34:0.3)
   		\centerarc[black,very thin](v2)(-18.43:29.05:0.5)
   		\centerarc[black,very thin](v2)(-45:29.05:0.3)
   		\centerarc[black,very thin](v2)(231.34:315:0.5)   		
		
		\draw [fill=black] (v1) circle (2pt);
		\draw [fill=black] (v2) circle (2pt);
		\draw [fill=black] (v3) circle (2pt);
		\draw [fill=black] (v4) circle (2pt);
		\draw [fill=black] (v5) circle (2pt);
		\draw [fill=black] (v6) circle (2pt);		

		\node[left = 1pt of v1] (text_v1) {$1$};
		\node[above = 1pt of v2] (text_v2) {$2$};
		\node[above right = 1pt of v3] (text_v3) {$3$};
		\node[below left = 1pt of v4] (text_v4) {$4$};
		\node[below right = 1pt of v5] (text_v5) {$5$};		
		\node[below right = 1pt of v6] (text_v6) {$6$};
	
		\end{tikzpicture}
		\caption{$w = u$} \label{fig:EL_example_new_assum_b}
	\end{subfigure}
	\caption{Examples where Assumption \ref{assum:new} is violated and satisfied, respectively.}
	\label{fig:EL_example_new_assum}
\end{figure}

Assumption \ref{assum:new} is illustrated in Figure \ref{fig:EL_example_new_assum}.
The figure provides two examples where Assumption \ref{assum:new} is satisfied and violated, respectively.
In Figure \ref{fig:EL_example_new_assum_a}, agent $2$ measures subtended angles between three pairs of other agents, i.e., $\alpha^{2}_{1,4}, \alpha^{2}_{4,1}, \alpha^{2}_{3,5}, \alpha^{2}_{5,3}, \alpha^{2}_{3,6}, \alpha^{2}_{6,3} \in \mathcal{S}$.
The bearing vector $\bold{g}_{1,2}$ has a geometric relationship with the bearing vector $\bold{g}_{4,2}$ by the subtended angle measurements $\alpha^{2}_{1,4}$ and $\alpha^{2}_{4,1}$.
It can be represented as $\angle{\bold{g}_{1,2}} = \angle{\bold{g}_{4,2}} + \alpha^{2}_{1,4} = \angle{\bold{g}_{4,2}} - \alpha^{2}_{4,1}$.
However, it can be seen that the bearing vector $\bold{g}_{1,2}$ does not have geometric relationships with the bearing vectors $\bold{g}_{3,2}$, $\bold{g}_{5,2}$, and $\bold{g}_{6,2}$ by the given subtended angle measurements.
In the case of Figure \ref{fig:EL_example_new_assum_a}, Assumption \ref{assum:new} is violated.
On the other hand, the example in Figure \ref{fig:EL_example_new_assum_b} satisfies Assumption \ref{assum:new}, since there are geometric relationships between all these bearing vectors.
For instance, they can be represented by the given subtended angle measurements as $\angle{\bold{g}_{1,2}} = \angle{\bold{g}_{3,2}} + \alpha^{2}_{1,4} + \alpha^{2}_{4,6} + \alpha^{2}_{6,3} = \angle{\bold{g}_{5,2}} + \alpha^{2}_{3,5} + \alpha^{2}_{1,4} + \alpha^{2}_{4,6} + \alpha^{2}_{6,3} = \angle{\bold{g}_{6,2}} + \alpha^{2}_{1,4} + \alpha^{2}_{4,6}$.
Assumption \ref{assum:new} guarantees that the given subtended angle measurements provide geometric relationships between all bearing vectors associated with them.

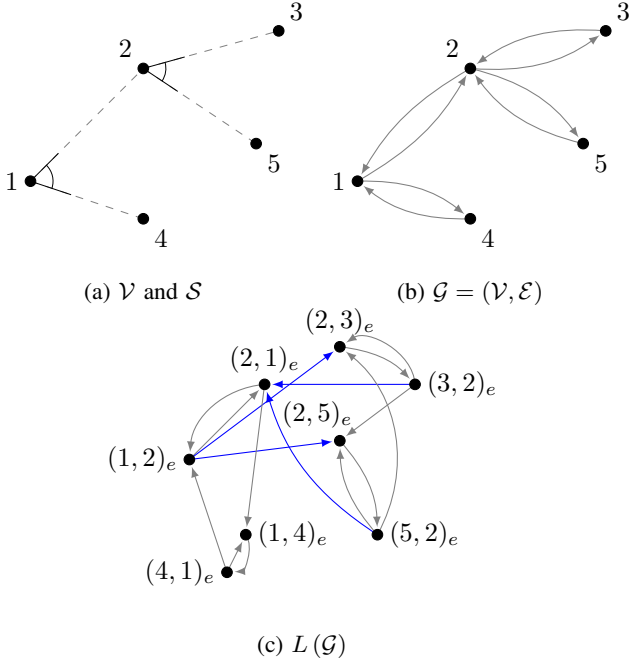
\begin{figure}[!h]
	\centering
	\begin{subfigure}{0.5\columnwidth}
		\centering
		\begin{tikzpicture}
		\useasboundingbox (0,0) rectangle (4,4);

		\coordinate (v1) at (0.5,1);
		\coordinate (v2) at (2,2.5);
		\coordinate (v3) at (3.8,3);		
		\coordinate (v4) at (2,0.5);
		\coordinate (v5) at (3.5,1.5);

		\draw[gray,very thin,dashed] (0.5,1) -- (2,2.5);
		\draw[gray,very thin,dashed] (0.5,1) -- (2,0.5);
		\draw[gray,very thin,dashed] (2,2.5) -- (3.8,3);
		\draw[gray,very thin,dashed] (2,2.5) -- (3.5,1.5);

		\node (e12) at ([shift=({45:0.7})]v1) {};
		\node (e14) at ([shift=({-18.43:0.7})]v1) {};
		\node (e23) at ([shift=({15.52:0.7})]v2) {};
		\node (e25) at ([shift=({-33.69:0.7})]v2) {};
		\draw[black,very thin] (v1) -- (e12);
		\draw[black,very thin] (v1) -- (e14);
		\draw[black,very thin] (v2) -- (e23);
		\draw[black,very thin] (v2) -- (e25);

   		\centerarc[black,very thin](v1)(-18.43:45:0.3)
   		\centerarc[black,very thin](v2)(-33.69:15.52:0.3)
		
		\draw [fill=black] (v1) circle (2pt);
		\draw [fill=black] (v2) circle (2pt);
		\draw [fill=black] (v3) circle (2pt);
		\draw [fill=black] (v4) circle (2pt);
		\draw [fill=black] (v5) circle (2pt);

		\node[left = 1pt of v1] (text_v1) {$1$};
		\node[above left= 1pt of v2] (text_v2) {$2$};
		\node[above right = 1pt of v3] (text_v3) {$3$};
		\node[below right = 1pt of v4] (text_v4) {$4$};
		\node[below right = 1pt of v5] (text_v5) {$5$};		
		
		\end{tikzpicture}
		\caption{$\mathcal{V}$ and $\mathcal{S}$} \label{fig:line_exp2_a}
	\end{subfigure}~
	\begin{subfigure}{0.5\columnwidth}
		\centering
		\begin{tikzpicture}
		\useasboundingbox (0,0) rectangle (4,4);

		\coordinate (v1) at (0.5,1);
		\coordinate (v2) at (2,2.5);
		\coordinate (v3) at (3.8,3);		
		\coordinate (v4) at (2,0.5);
		\coordinate (v5) at (3.5,1.5);

		\draw [gray, -latex, shorten >= 0.1cm] (v1) to[out=30,in=240] (v2);
		\draw [gray, -latex, shorten >= 0.1cm] (v2) to[out=210,in=60] (v1);		

		\draw [gray, -latex, shorten >= 0.1cm] (v1) to[out=1.57,in=141.57] (v4);	
		\draw [gray, -latex, shorten >= 0.1cm] (v4) to[out=181.57,in=-38.43] (v1);	

		\draw [gray, -latex, shorten >= 0.1cm] (v2) to[out=-4.48,in=215.52] (v3);
		\draw [gray, -latex, shorten >= 0.1cm] (v3) to[out=175.52,in=35.52] (v2);

		\draw [gray, -latex, shorten >= 0.1cm] (v2) to[out=-13.69,in=126.31] (v5);				
		\draw [gray, -latex, shorten >= 0.1cm] (v5) to[out=166.31,in=-53.69] (v2);

		\draw [fill=black] (v1) circle (2pt);
		\draw [fill=black] (v2) circle (2pt);
		\draw [fill=black] (v3) circle (2pt);
		\draw [fill=black] (v4) circle (2pt);
		\draw [fill=black] (v5) circle (2pt);

		\node[left = 1pt of v1] (text_v1) {$1$};
		\node[above left= 1pt of v2] (text_v2) {$2$};
		\node[above right = 1pt of v3] (text_v3) {$3$};
		\node[below right = 1pt of v4] (text_v4) {$4$};
		\node[below right = 1pt of v5] (text_v5) {$5$};		
		
		\end{tikzpicture}
		\caption{$\mathcal{G} = \left( \mathcal{V}, \mathcal{E} \right)$} \label{fig:line_exp2_b}
	\end{subfigure}\\
	\begin{subfigure}{0.5\columnwidth}
		\centering
		\begin{tikzpicture}
		\useasboundingbox (0,0) rectangle (4,4);

		\coordinate (e12) at (0.5,2);
		\coordinate (e21) at (1.5,3);

		\coordinate (e23) at (2.5,3.5);
		\coordinate (e32) at (3.5,3);		
		
		\coordinate (e14) at (1.25,1);
		\coordinate (e41) at (1,0.5);

		\coordinate (e25) at (2.5,2.25);
		\coordinate (e52) at (3,1);

		\draw [gray, -latex, shorten >= 0.1cm] (e12) -- (e21);
		\draw [blue, -latex, shorten >= 0.1cm] (e12) -- (e23);
		\draw [blue, -latex, shorten >= 0.1cm] (e12) -- (e25);

		\draw [gray, -latex, shorten >= 0.1cm] (e21) -- (e14);
		\draw [gray, -latex, shorten >= 0.1cm] (e21) to[out=185,in=85] (e12);
		
		\draw [gray, -latex, shorten >= 0.1cm] (e41) -- (e14);
		\draw [gray, -latex, shorten >= 0.1cm] (e41) -- (e12);
		
		\draw [gray, -latex, shorten >= 0.1cm] (e14) to[out=303.43,in=3.43] (e41);
		
		\draw [gray, -latex, shorten >= 0.1cm] (e23) to[out=-6.57,in=133.43] (e32);
		
		\draw [gray, -latex, shorten >= 0.1cm] (e32) -- (e25);		
		\draw [gray, -latex, shorten >= 0.1cm] (e32) to[out=93.43,in=33.43] (e23);
		\draw [blue, -latex, shorten >= 0.1cm] (e32) -- (e21);		
		
		\draw [gray, -latex, shorten >= 0.1cm] (e25) to[out=-48.2,in=91.8] (e52);

		\draw [gray, -latex, shorten >= 0.1cm] (e52) to[out=61.31,in=-38.69] (e23);
		\draw [gray, -latex, shorten >= 0.1cm] (e52) to[out=131.8,in=-88.2] (e25);
		\draw [blue, -latex, shorten >= 0.1cm] (e52) to[out=146.87,in=-73.13] (e21);

		\draw [fill=black] (e12) circle (2pt);
		\draw [fill=black] (e21) circle (2pt);

		\draw [fill=black] (e23) circle (2pt);
		\draw [fill=black] (e32) circle (2pt);		

		\draw [fill=black] (e14) circle (2pt);
		\draw [fill=black] (e41) circle (2pt);
		
		\draw [fill=black] (e25) circle (2pt);
		\draw [fill=black] (e52) circle (2pt);

		\node[left = 1pt of e12] (text_e12) {$( 1,2 )_{e}$};
		\node[above = 1pt of e21] (text_e21) {$( 2,1 )_{e}$};
		
		\node[above = 1pt of e23] (text_e23) {$( 2,3 )_{e}$};
		\node[right = 1pt of e32] (text_e23) {$( 3,2 )_{e}$};				

		\node[right = 1pt of e14] (text_e14) {$( 1,4 )_{e}$};
		\node[left = 1pt of e41] (text_e41) {$( 4,1 )_{e}$};

		\node (text_e25) at (2.2,2.6)  {$( 2,5 )_{e}$};
		\node[right = 1pt of e52] (text_e52) {$( 5,2 )_{e}$};

		\end{tikzpicture}
		\caption{$L\left(\mathcal{G}\right)$} \label{fig:line_exp2_c}
	\end{subfigure}
	\caption{The communication graph $\mathcal{G}$ and its directed line graph $L\left(\mathcal{G}\right)$ in Example \ref{exp:1}(a).}	
	\label{fig:line_exp2}
\end{figure}

Although the concept of directed line graph is an appropriate approach to define the edge agents, before applying it, $\mathcal{G}$ needs to be further modified to overcome possible undesired connections, as explained next.
\begin{definition}[Undesired connection] \label{def:unde_conn}
For the given communication graph $\mathcal{G}$ and set $\mathcal{S}$ of subtended angle measurements, an edge in $L\left( \mathcal{G} \right)$ from edge node $\left( u,v \right)_{e} \in \mathcal{E}$ to edge node $\left( v,w \right)_{e} \in \mathcal{E}$ is called undesired connection if $u \neq w$ and there does not exist $\alpha^{v}_{wu} \in \mathcal{S}$.
\end{definition}
Definition \ref{def:unde_conn} is illustrated in Example \ref{exp:1}(a) and Figure \ref{fig:line_exp2}.	
\begin{example}[a]\label{exp:1}
Let a communication graph $\mathcal{G}=\left( \mathcal{V}, \mathcal{E} \right)$ with $\mathcal{V} = \left\{ 1, 2, 3, 4, 5 \right\}$ and $\mathcal{S} = \left\{ \alpha^{1}_{2,4}, \alpha^{1}_{4,2}, \alpha^{2}_{3,5}, \alpha^{2}_{5,3} \right\}$, be given.
The communication graph $\mathcal{G}$ and its directed line graph $L\left(\mathcal{G}\right)$ are as shown in Figure \ref{fig:line_exp2}. 
In Figure \ref{fig:line_exp2_a}, there are five agents and two couples of subtended angles are measured.
By Assumption \ref{assum:1}, the communication graph $\mathcal{G}$ is constructed as illustrated in Figure \ref{fig:line_exp2_b},
and its directed line graph $L\left(\mathcal{G}\right)$ is shown in Figure \ref{fig:line_exp2_c}.
In this example, it can be seen that the edge nodes $\left( 2,3 \right)_{e}$ and $\left( 2,5 \right)_{e}$ are neighbors of the edge node $\left( 1,2 \right)_{e}$.
Let $\left( u,v \right)^{\epsilon}$ be an edge agent corresponding to an edge node $\left( u,v \right)_{e}$ and assume that $L\left(\mathcal{G}\right)$ is used as the interaction graph of edge agents.
The edge agent $\left( 1,2 \right)^{\epsilon}$ has a neighbor $\left( 2,3 \right)^{\epsilon}$.
However, the subtended angle measurements $\alpha^{2}_{1,3}$ and $\alpha^{2}_{3,1}$ are not available.
Thus, the geometric relationships (\ref{eq:geo_re}) cannot be used for $\left( u,v \right) = \left( 1,2 \right)$ and $\left( v,w \right) = \left( 2,3 \right)$.
In other words, the edge agent $\left( 1,2 \right)^{\epsilon}$ can not use the orientation angle of its neighbor, $\angle{\bold{g}_{3,2}}$, to estimate its own orientation $\angle{\bold{g}_{2,1}}$.
Similarly, the orientation information of the edge agent $\left( 2,5 \right)^{\epsilon}$ is not available to the edge agent $\left( 1,2 \right)^{\epsilon}$.
This means that the edge node $\left( 1,2 \right)_{e}$ is not supposed to be connected to the edge nodes $\left( 2,3 \right)_{e}$ and $\left( 2,5 \right)_{e}$ in $L\left( \mathcal{G} \right)$.
For the same reason, the edge nodes $\left( 3,2 \right)_{e}$ and $\left( 5,2 \right)_{e}$ are not supposed to be connected to the edge node $\left( 2,1 \right)_{e}$.
In Figure \ref{fig:line_exp2_c}, the blue lines are the undesired connections.
\end{example}

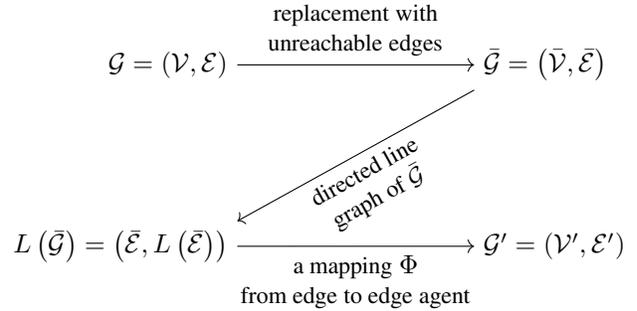
\begin{figure}[h]
\begin{tikzpicture}
  \matrix (m) [matrix of math nodes, row sep=5em, column sep=9em, minimum width=2em, column 1/.style={anchor=base east}, column 2/.style={anchor=base west}]
  {
     \mathcal{G} = \left( \mathcal{V}, \mathcal{E} \right) & \bar{\mathcal{G}} = \left( \bar{\mathcal{V}}, \bar{\mathcal{E}} \right) \\
     L\left(\bar{\mathcal{G}}\right) = \left( \bar{\mathcal{E}}, L\left(\bar{\mathcal{E}}\right) \right) & \mathcal{G}' = \left( \mathcal{V}', \mathcal{E}' \right) \\};
	
	\draw[->] (m-1-1) -- (m-1-2) node[midway, above, align=center] {\small replacement with\\ \small unreachable edges};
	\draw[->] (m-1-2.south west) -- (m-2-1.north east) node[midway, below, align=center, sloped] {\small directed line \\ \small graph of $\bar{\mathcal{G}}$};
	\draw[->] (m-2-1) -- (m-2-2) node[midway, below, align=center] {\small a mapping \normalsize $\Phi$ \\ \small from edge to edge agent};
\end{tikzpicture}
\caption{Relation of the communication graph $\mathcal{G}$, the edge localization graph $\bar{\mathcal{G}}$, and the localization interaction graph $\mathcal{G}'$.}
\label{fig:graph_relation}
\end{figure}

In this paper, the undesired connections are overcome by adoption of virtual vertices, called \textit{unreachable virtual vertices} (to be formally defined in the sequel)
and the communication graph $\mathcal{G}$ is converted into an \textit{edge localization graph}, $\bar{\mathcal{G}} = \left( \bar{\mathcal{V}}, \bar{\mathcal{E}}\right)$, where the unreachable virtual vertices are added to $\mathcal{V}$ to form $\bar{\mathcal{V}} \supseteq \mathcal{V}$.
The directed line graph $L \left( \bar{\mathcal{G}} \right)$ of the edge localization graph is used to represent the interactions between edge agents.
Finally, an interaction graph of the edge agents, called \textit{localization interaction graph}, is constructed by the directed line graph through a mapping.
Figure \ref{fig:graph_relation} depicts the relation among the communication graph $\mathcal{G}$, the edge localization graph $\bar{\mathcal{G}}$, and the localization interaction graph $\mathcal{G}'$.
The procedure and the terms above are formally presented in the remainder of this section.

In the aforementioned modification of $\mathcal{G}$ to overcome undesired connections, the following notion is utilized.
\begin{definition}[Unreachable virtual vertex] \label{def:un_re_v}
Given an edge $\left( u,v \right) \in \mathcal{E}$, if there does not exist $w$ such that $\alpha^{v}_{wu} \in \mathcal{S}$ exists, then the vertex $v$ is replaced by unreachable virtual vertex\footnote{Even for a same vertex, there could be multiple different unreachable virtual vertices.} $\bar{v}^u$, which is co-located with $v$.
\end{definition}
The replacements of a real vertex by unreachable virtual vertices do not affect the orientation of edge agent, since the real vertex and the unreachable virtual vertices are co-located.
The directed edge, whose head or tail is replaced with an unreachable virtual vertex, is called unreachable edge in this paper.
Note that an unreachable edge whose head and tail are both unreachable virtual vertices does not exist, because there must exist at least a subtended angle measured at its head or tail by Assumption \ref{assum:1}. 
To investigate the edge localizability, let us first define edge localization graph.
\begin{definition}[Edge localization graph] \label{def:re_g}
An edge localization graph $\bar{\mathcal{G}} = \left( \bar{\mathcal{V}}, \bar{\mathcal{E}} \right)$ of the communication graph $\mathcal{G} = \left( \mathcal{V}, \mathcal{E} \right)$ is a directed graph whose vertex set and edge set are defined as follows:
\begin{align*}
\bar{\mathcal{V}} =& \left(\mathcal{V} \setminus \left\{ u \mid u \in \mathcal{V},\; \nexists \alpha^{u}_{wv} \in \mathcal{S} \; \forall v,w \in \mathcal{V} \right\} \right) \cup \mathcal{V^u} \\
\bar{\mathcal{E}} =& \left( \mathcal{E} \setminus \left\{ \left(u, v\right), \left(v, u\right) \mid \left( u,v \right) \in \mathcal{E},\; \nexists w \in \mathcal{V} \;{\rm{s.t.}}\; \alpha^{v}_{wu} \in \mathcal{S} \right\} \right) 	\\
&\cup \mathcal{E^u}
\end{align*}
where $\mathcal{E^u} = \bigl\{ \left(u, \bar{v}^{u}\right),\; \left(\bar{v}^{u}, u\right) \mid \left( u,v \right) \in \mathcal{E},\; \nexists w \in \mathcal{V} \;{\rm{s.t.}}$ $\alpha^{v}_{wu} \in \mathcal{S} \bigr\}$ and $\mathcal{V^u} = \bigl\{ \bar{u}^{v} \mid \left(\bar{u}^{v}, v\right) \in \mathcal{E^u} \bigr\}$.
\end{definition}
For a pair of edges $\left(u,v\right)$, $\left(v,u\right) \in \mathcal{E}$, they are replaced with the unreachable edges $\left(u,\bar{v}^{u}\right)$ and $\left(\bar{v}^{u},u\right)$, respectively, in the edge localization graph $\bar{\mathcal{G}}$ if the agent $v$ does not measure a subtended angle between the agent $u$ and any other agent.
The agents, which do not measure any subtended angle, in the communication graph $\mathcal{G}$ are removed from the vertex set of the edge localization graph $\bar{\mathcal{G}}$.
Here we give an example to show the edge localization graph.
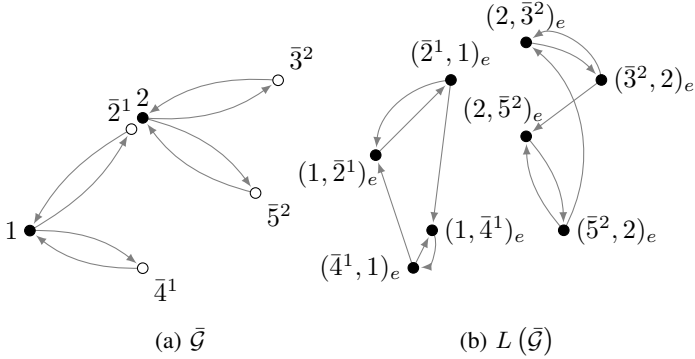
\begin{figure}[!h]
	\centering
	\begin{subfigure}{0.5\columnwidth}
		\centering
		\begin{tikzpicture}
		\useasboundingbox (0.5,0) rectangle (4.5,4);

		\coordinate (v1) at (0.5,1);
		\coordinate (v2) at (2,2.5);
		\coordinate (v3) at (3.8,3);		
		\coordinate (v4) at (2,0.5);
		\coordinate (v5) at (3.5,1.5);

		\coordinate (v2u) at ($(v1)!0.9!(v2)$) {};

		\draw [gray, -latex, shorten >= 0.1cm] (v1) to[out=30,in=240] (v2u);
		\draw [gray, -latex, shorten >= 0.1cm] (v2u) to[out=210,in=60] (v1);		

		\draw [gray, -latex, shorten >= 0.1cm] (v1) to[out=1.57,in=141.57] (v4);	
		\draw [gray, -latex, shorten >= 0.1cm] (v4) to[out=181.57,in=-38.43] (v1);	

		\draw [gray, -latex, shorten >= 0.1cm] (v2) to[out=-4.48,in=215.52] (v3);
		\draw [gray, -latex, shorten >= 0.1cm] (v3) to[out=175.52,in=35.52] (v2);

		\draw [gray, -latex, shorten >= 0.1cm] (v2) to[out=-13.69,in=126.31] (v5);				
		\draw [gray, -latex, shorten >= 0.1cm] (v5) to[out=166.31,in=-53.69] (v2);

		\draw [fill=black] (v1) circle (2pt);
		\draw [fill=black] (v2) circle (2pt);
		\draw [fill=white] (v3) circle (2pt);
		\draw [fill=white] (v4) circle (2pt);
		\draw [fill=white] (v5) circle (2pt);
		
		\draw [fill=white] (v2u) circle (2pt);

		\node[left = 1pt of v1] (text_v1) {$1$};
		\node[above = 1pt of v2] (text_v2) {$2$};
		\node[above right = 1pt of v3] (text_v3) {$\bar{3}^2$};
		\node[below right = 1pt of v4] (text_v4) {$\bar{4}^1$};
		\node[below right = 1pt of v5] (text_v5) {$\bar{5}^2$};		

		\node (text_v2u) at (1.7,2.6) {$\bar{2}^1$};
				
		\end{tikzpicture}
		\caption{$\bar{\mathcal{G}}$}\label{fig:line_exp3_a}
	\end{subfigure}~
	\begin{subfigure}{0.5\columnwidth}
		\centering
		\begin{tikzpicture}
		\useasboundingbox (0.5,0) rectangle (4,4);

		\coordinate (e12) at (0.5,2);
		\coordinate (e21) at (1.5,3);

		\coordinate (e23) at (2.5,3.5);
		\coordinate (e32) at (3.5,3);		
		
		\coordinate (e14) at (1.25,1);
		\coordinate (e41) at (1,0.5);

		\coordinate (e25) at (2.5,2.25);
		\coordinate (e52) at (3,1);

		\draw [gray, -latex, shorten >= 0.1cm] (e12) -- (e21);

		\draw [gray, -latex, shorten >= 0.1cm] (e21) -- (e14);
		\draw [gray, -latex, shorten >= 0.1cm] (e21) to[out=185,in=85] (e12);
		
		\draw [gray, -latex, shorten >= 0.1cm] (e41) -- (e14);
		\draw [gray, -latex, shorten >= 0.1cm] (e41) -- (e12);
		
		\draw [gray, -latex, shorten >= 0.1cm] (e14) to[out=303.43,in=3.43] (e41);
		
		\draw [gray, -latex, shorten >= 0.1cm] (e23) to[out=-6.57,in=133.43] (e32);
		
		\draw [gray, -latex, shorten >= 0.1cm] (e32) -- (e25);		
		\draw [gray, -latex, shorten >= 0.1cm] (e32) to[out=93.43,in=33.43] (e23);
		
		\draw [gray, -latex, shorten >= 0.1cm] (e25) to[out=-48.2,in=91.8] (e52);

		\draw [gray, -latex, shorten >= 0.1cm] (e52) to[out=61.31,in=-38.69] (e23);
		\draw [gray, -latex, shorten >= 0.1cm] (e52) to[out=131.8,in=-88.2] (e25);

		\draw [fill=black] (e12) circle (2pt);
		\draw [fill=black] (e21) circle (2pt);

		\draw [fill=black] (e23) circle (2pt);
		\draw [fill=black] (e32) circle (2pt);		

		\draw [fill=black] (e14) circle (2pt);
		\draw [fill=black] (e41) circle (2pt);
		
		\draw [fill=black] (e25) circle (2pt);
		\draw [fill=black] (e52) circle (2pt);

		\node (text_e12) at (0,1.75) {$( 1,\bar{2}^1 )_{e}$};
		\node[above = 1pt of e21] (text_e21) {$( \bar{2}^1,1 )_{e}$};
		
		\node[above = 1pt of e23] (text_e23) {$( 2,\bar{3}^2 )_{e}$};
		\node[right = 1pt of e32] (text_e23) {$( \bar{3}^2,2 )_{e}$};				

		\node[right = 1pt of e14] (text_e14) {$( 1,\bar{4}^1 )_{e}$};
		\node[left = 1pt of e41] (text_e41) {$( \bar{4}^1,1 )_{e}$};

		\node (text_e25) at (2.2,2.6)  {$( 2,\bar{5}^2 )_{e}$};
		\node[right = 1pt of e52] (text_e52) {$( \bar{5}^2,2 )_{e}$};

		\end{tikzpicture}
		\caption{$L\left(\bar{\mathcal{G}}\right)$} \label{fig:line_exp3_b}
	\end{subfigure}
	\caption{The edge localization graph $\bar{\mathcal{G}}$ and its directed line graph $L\left( \bar{\mathcal{G}} \right)$ in Example \ref{exp:2}(b).}	
	\label{fig:line_exp3}
\end{figure}
\setcounter{example}{0}
\begin{example}[b]\label{exp:2}
Let $\mathcal{V}$ and $\mathcal{S}$ be given same as in Example \ref{exp:1}(a).		
Then the edge localization graph $\bar{\mathcal{G}}$ and its directed line graph $L\left( \bar{\mathcal{G}} \right)$ are given as illustrated in Figure \ref{fig:line_exp3}.
In Figure \ref{fig:line_exp3_a}, empty dots represent the unreachable virtual vertex different from the real vertex.
As shown in Figure \ref{fig:line_exp3_b}, the undesired connections in Example \ref{exp:1}(a) are removed in the the directed line graph of the edge localization graph, $L\left( \bar{\mathcal{G}} \right)$.	
Let $\left( 1,2 \right)^{\epsilon}$, $\left( 2,1 \right)^{\epsilon}$, $\left( 2,3 \right)^{\epsilon}$, $\left( 3,2 \right)^{\epsilon}$, $\left( 2,5 \right)^{\epsilon}$, and $\left( 5,2 \right)^{\epsilon}$ be the edge agents corresponding to the edge nodes $( 1,\bar{2}^1 )_{e}$, $( \bar{2}^1,1 )_{e}$, $( 2,\bar{3}^2 )_{e}$, $( \bar{3}^2,2 )_{e}$, $( 2,\bar{5}^2 )_{e}$, and $( \bar{5}^2,2 )_{e}$, respectively.
Assume that $L\left( \bar{\mathcal{G}} \right)$ is used as the interaction graph of edge agents.
Then the edge agent $\left( 1,2 \right)^{\epsilon}$ is not connected to the edge agents $\left( 2,3 \right)^{\epsilon}$ and $\left( 2,5 \right)^{\epsilon}$.
The edge agents $\left( 3,2 \right)^{\epsilon}$ and $\left( 5,2 \right)^{\epsilon}$ also do not reach the edge agent $\left( 2,1 \right)^{\epsilon}$.
These mean that each edge agent $\left( u,v \right)^{\epsilon}$ does not have a neighbor $\left( v,w \right)^{\epsilon}$ if the orientation of $\left( u,v \right)^{\epsilon}$ can not be calculated from the orientation angle of $\left( v,w \right)^{\epsilon}$ through the geometric relationships (\ref{eq:geo_re}).
\end{example}

\begin{figure}[!h]	
	\centering
	\begin{subfigure}{0.5\columnwidth}
		\centering
		\begin{tikzpicture}		
		\useasboundingbox (0,0) rectangle (4,4);

		\coordinate (v1) at (0.5,0.5);
		\coordinate (v2) at (0.5,3.5);
		\coordinate (v3) at (3.5,3.5);		
		\coordinate (v4) at (3.5,0.5);

		\draw[gray,very thin,dashed] (v1) -- (v2);
		\draw[gray,very thin,dashed] (v1) -- (v4);		
		\draw[gray,very thin,dashed] (v2) -- (v3);		
		\draw[gray,very thin,dashed] (v2) -- (v4);

		\node (e12) at ([shift=({90:1.1})]v1) {};
		\node (e14) at ([shift=({0:1.1})]v1) {};
		\node (e23) at ([shift=({0:1.1})]v2) {};
		\node (e24) at ([shift=({-45:1.1})]v2) {};
		\node (e41) at ([shift=({180:1.1})]v4) {};
		\node (e42) at ([shift=({135:1.1})]v4) {};
		
		\draw[black,very thin] (v1) -- (e12);
		\draw[black,very thin] (v1) -- (e14);		
		\draw[black,very thin] (v2) -- (e23);				
		\draw[black,very thin] (v2) -- (e24);
		\draw[black,very thin] (v4) -- (e41);
		\draw[black,very thin] (v4) -- (e42);				
		
   		\centerarc[black,very thin](v1)(0:90:0.3)				
   		\centerarc[black,very thin](v2)(0:-45:0.3)
   		\centerarc[black,very thin](v4)(135:180:0.3)

		\draw [fill=black] (v1) circle (2pt);
		\draw [fill=black] (v2) circle (2pt);
		\draw [fill=black] (v3) circle (2pt);
		\draw [fill=black] (v4) circle (2pt);

		\node (text_v1) at (0.5-0.2,0.5-0.2) {$1$};
		\node (text_v2) at (0.5-0.2,3.5+0.2) {$2$};
		\node (text_v3) at (3.5+0.2,3.5+0.2) {$3$};
		\node (text_v4) at (3.5+0.2,0.5-0.2) {$4$};
		
		\node (text_a1) at (0.5+0.75,0.5+0.5) {\footnotesize{$\alpha^{1}_{2,4}, \alpha^{1}_{4,2}$}};		
		\node (text_a4) at (3.5-0.95,0.5+0.25) {\footnotesize{$\alpha^{4}_{1,2}, \alpha^{4}_{2,1}$}};
		\node (text_a2) at (0.5+1,3.5-0.25) {\footnotesize{$\alpha^{2}_{3,4}, \alpha^{2}_{4,3}$}};
				
		\end{tikzpicture}
		\caption{$\mathcal{V}$ and $\mathcal{S}$} \label{fig:pro_exp_a}
	\end{subfigure}~
	\begin{subfigure}{0.5\columnwidth}
		\centering
		\begin{tikzpicture}		
		\useasboundingbox (0,0) rectangle (4,4);

		\coordinate (v1) at (0.5,0.5);
		\coordinate (v2) at (0.5,3.5);
		\coordinate (v3) at (3.5,3.5);		
		\coordinate (v4) at (3.5,0.5);

		\draw [gray, -latex, shorten >= 0.1cm] (v1) to[out=90+10,in=-90-10] (v2);
		\draw [gray, -latex, shorten >= 0.1cm] (v2) to[out=-90+10,in=90-10] (v1);		
		\draw [gray, -latex, shorten >= 0.1cm] (v1) to[out=0-10,in=180+10] (v4);
		\draw [gray, -latex, shorten >= 0.1cm] (v4) to[out=180-10,in=0+10] (v1);		
		\draw [gray, -latex, shorten >= 0.1cm] (v2) to[out=0-10,in=180+10] (v3);
		\draw [gray, -latex, shorten >= 0.1cm] (v3) to[out=180-10,in=0+10] (v2);		
		\draw [gray, -latex, shorten >= 0.1cm] (v2) to[out=-45+10,in=135-10] (v4);
		\draw [gray, -latex, shorten >= 0.1cm] (v4) to[out=135+10,in=-45-10] (v2);		

		\draw [fill=black] (v1) circle (2pt);
		\draw [fill=black] (v2) circle (2pt);
		\draw [fill=black] (v3) circle (2pt);
		\draw [fill=black] (v4) circle (2pt);

		\node (text_v1) at (0.5-0.2,0.5-0.2) {$1$};
		\node (text_v2) at (0.5-0.2,3.5+0.2) {$2$};
		\node (text_v3) at (3.5+0.2,3.5+0.2) {$3$};
		\node (text_v4) at (3.5+0.2,0.5-0.2) {$4$};

		\node (text_e12) at (0.5-0.5,2) {\tiny{$\left( 1,2 \right)$}};
		\node (text_e21) at (0.5+0.5,2) {\tiny{$\left( 2,1 \right)$}};		
		\node (text_e14) at (2,0.5-0.3) {\tiny{$\left( 1,4 \right)$}};
		\node (text_e41) at (2,0.5+0.3) {\tiny{$\left( 4,1 \right)$}};		
		\node (text_e23) at (2,3.5-0.3) {\tiny{$\left( 2,3 \right)$}};
		\node (text_e32) at (2,3.5+0.3) {\tiny{$\left( 3,2 \right)$}};
		\node (text_e24) at (2+0.37,2+0.37) {\tiny{$\left( 2,4 \right)$}};
		\node (text_e42) at (2-0.37,2-0.37) {\tiny{$\left( 4,2 \right)$}};

		\end{tikzpicture}
		\caption{$\mathcal{G}$} \label{fig:pro_exp_b}
	\end{subfigure}\\	\vspace{0.6cm}
	\begin{subfigure}{0.5\columnwidth}
		\centering
		\begin{tikzpicture}		
		\useasboundingbox (0,0) rectangle (4,4);

		\coordinate (v1) at (0.5,0.5);
		\coordinate (v2) at (0.5,3.5);
		\coordinate (v3) at (3.5,3.5);		
		\coordinate (v4) at (3.5,0.5);

		\coordinate (v2u) at ($(v1)!0.9!(v2)$) {};

		\draw [gray, -latex, shorten >= 0.1cm] (v1) to[out=90+10,in=-90-10] (v2u);
		\draw [gray, -latex, shorten >= 0.1cm] (v2u) to[out=-90+10,in=90-10] (v1);		
		\draw [gray, -latex, shorten >= 0.1cm] (v1) to[out=0-10,in=180+10] (v4);
		\draw [gray, -latex, shorten >= 0.1cm] (v4) to[out=180-10,in=0+10] (v1);		
		\draw [gray, -latex, shorten >= 0.1cm] (v2) to[out=0-10,in=180+10] (v3);
		\draw [gray, -latex, shorten >= 0.1cm] (v3) to[out=180-10,in=0+10] (v2);		
		\draw [gray, -latex, shorten >= 0.1cm] (v2) to[out=-45+10,in=135-10] (v4);
		\draw [gray, -latex, shorten >= 0.1cm] (v4) to[out=135+10,in=-45-10] (v2);		

		\draw [fill=black] (v1) circle (2pt);
		\draw [fill=black] (v2) circle (2pt);
		\draw [fill=white] (v2u) circle (2pt);		
		\draw [fill=white] (v3) circle (2pt);
		\draw [fill=black] (v4) circle (2pt);

		\node (text_v1) at (0.5-0.2,0.5-0.2) {$1$};
		\node (text_v2) at (0.5-0.2,3.5+0.2) {$2$};
		\node (text_v3) at (3.5+0.25,3.5+0.25) {$\bar{3}^2$};
		\node (text_v4) at (3.5+0.2,0.5-0.2) {$4$};
		\node (text_v2) at (0.5-0.2,3.5-0.2) {$\bar{2}^1$};

		\node (text_e12) at (0.5-0.55,2-0.10) {\tiny{$( 1,\bar{2}^1 )$}};
		\node (text_e21) at (0.5+0.55,2-0.10) {\tiny{$( \bar{2}^1,1 )$}};		
		\node (text_e14) at (2,0.5-0.3) {\tiny{$\left( 1,4 \right)$}};
		\node (text_e41) at (2,0.5+0.3) {\tiny{$\left( 4,1 \right)$}};		
		\node (text_e23) at (2,3.5-0.35) {\tiny{$( 2,\bar{3}^2 )$}};
		\node (text_e32) at (2,3.5+0.35) {\tiny{$( \bar{3}^2,2 )$}};
		\node (text_e24) at (2+0.37,2+0.37) {\tiny{$\left( 2,4 \right)$}};
		\node (text_e42) at (2-0.37,2-0.37) {\tiny{$\left( 4,2 \right)$}};
	
		\end{tikzpicture}
		\caption{$\bar{\mathcal{G}}$} \label{fig:pro_exp_c}
	\end{subfigure}~
	\begin{subfigure}{0.5\columnwidth}
		\centering
		\begin{tikzpicture}		
		\useasboundingbox (0,0) rectangle (4,4);


		\coordinate (e12) at (0.5,1.5);
		\coordinate (e14) at (1.5,0.5);
		\coordinate (e21) at (0.5,2.5);
		\coordinate (e23) at (1.5,3.5);
		\coordinate (e24) at (1.5,2.5);		
		\coordinate (e32) at (3.5,3.5);
		\coordinate (e41) at (3.0,0.5);		
		\coordinate (e42) at (3.0,1.5);

		\draw [gray, -latex, shorten >= 0.1cm] (e12) to[out=90+30,in=-90-30] (e21);

		\draw [gray, -latex, shorten >= 0.1cm] (e14) to[out=0-30,in=180+30] (e41);
		\draw [gray, -latex, shorten >= 0.1cm] (e14) -- (e42);

		\draw [gray, -latex, shorten >= 0.1cm] (e21) -- (e12);
		\draw [gray, -latex, shorten >= 0.1cm] (e21) -- (e14);		
		
		\draw [gray, -latex, shorten >= 0.1cm] (e23) to[out=0+30,in=180-30] (e32);
		
		\draw [gray, -latex, shorten >= 0.1cm] (e24) to[out=-55.01-15,in=124.99+15] (e41);
		\draw [gray, -latex, shorten >= 0.1cm] (e24) to[out=-45-20,in=135+20] (e42);				
		\draw [gray, -latex, shorten >= 0.1cm] (e32) -- (e23);
		\draw [gray, -latex, shorten >= 0.1cm] (e32) -- (e24);

		\draw [gray, -latex, shorten >= 0.1cm] (e41) -- (e12);
		\draw [gray, -latex, shorten >= 0.1cm] (e41) -- (e14);
		
		\draw [gray, -latex, shorten >= 0.1cm] (e42) to[out=135-15,in=-45+15] (e24);			
		\draw [gray, -latex, shorten >= 0.1cm] (e42) to[out=122.47-20,in=-57.53+20] (e23);

		\draw [fill=black] (e12) circle (2pt);
		\draw [fill=black] (e14) circle (2pt);
		\draw [fill=black] (e21) circle (2pt);
		\draw [fill=black] (e23) circle (2pt);
		\draw [fill=black] (e24) circle (2pt);		
		\draw [fill=black] (e32) circle (2pt);
		\draw [fill=black] (e41) circle (2pt);
		\draw [fill=black] (e42) circle (2pt);		

		\node[below = 1.7pt of e12] (text_e12) {\scriptsize{$( 1,\bar{2}^{1} )_{e}$}};
		\node[left = 1pt of e14] (text_e12) {\scriptsize{$( 1,4 )_{e}$}};
		\node[above = 1pt of e21] (text_e21) {\scriptsize{$( \bar{2}^{1},1 )_{e}$}};
		\node[left = 1pt of e23] (text_e23) {\scriptsize{$( 2,\bar{3}^{2} )_{e}$}};
		\node[above = 1pt of e24] (text_e24) {\scriptsize{$( 2,4 )_{e}$}};
		\node[below = 1.7pt of e32] (text_e32) {\scriptsize{$( \bar{3}^{2},2 )_{e}$}};
		\node[right = 1pt of e41] (text_e41) {\scriptsize{$( 4,1 )_{e}$}};
		\node[right = 1pt of e42] (text_e42) {\scriptsize{$( 4,2 )_{e}$}};

		\end{tikzpicture}
		\caption{$L\left(\bar{\mathcal{G}}\right)$} \label{fig:pro_exp_d}
	\end{subfigure}\\	\vspace{0.6cm}
	\begin{subfigure}{0.5\columnwidth}
		\centering
		\begin{tikzpicture}		
		\useasboundingbox (0,0) rectangle (4,4);


		\coordinate (e12) at (0.5,1.5);
		\coordinate (e14) at (1.5,0.5);
		\coordinate (e21) at (0.5,2.5);
		\coordinate (e23) at (1.5,3.5);
		\coordinate (e24) at (1.5,2.5);		
		\coordinate (e32) at (3.5,3.5);
		\coordinate (e41) at (3.0,0.5);		
		\coordinate (e42) at (3.0,1.5);

		\draw [gray, -latex, shorten >= 0.1cm] (e12) to[out=90+30,in=-90-30] (e21);

		\draw [gray, -latex, shorten >= 0.1cm] (e14) to[out=0-30,in=180+30] (e41);
		\draw [gray, -latex, shorten >= 0.1cm] (e14) -- (e42);

		\draw [gray, -latex, shorten >= 0.1cm] (e21) -- (e12);
		\draw [gray, -latex, shorten >= 0.1cm] (e21) -- (e14);		
		
		\draw [gray, -latex, shorten >= 0.1cm] (e23) to[out=0+30,in=180-30] (e32);
		
		\draw [gray, -latex, shorten >= 0.1cm] (e24) to[out=-55.01-15,in=124.99+15] (e41);
		\draw [gray, -latex, shorten >= 0.1cm] (e24) to[out=-45-20,in=135+20] (e42);				
		\draw [gray, -latex, shorten >= 0.1cm] (e32) -- (e23);
		\draw [gray, -latex, shorten >= 0.1cm] (e32) -- (e24);

		\draw [gray, -latex, shorten >= 0.1cm] (e41) -- (e12);
		\draw [gray, -latex, shorten >= 0.1cm] (e41) -- (e14);
		
		\draw [gray, -latex, shorten >= 0.1cm] (e42) to[out=135-15,in=-45+15] (e24);			
		\draw [gray, -latex, shorten >= 0.1cm] (e42) to[out=122.47-20,in=-57.53+20] (e23);

		\draw [fill=black] (e12) circle (2pt);
		\draw [fill=black] (e14) circle (2pt);
		\draw [fill=black] (e21) circle (2pt);
		\draw [fill=black] (e23) circle (2pt);
		\draw [fill=black] (e24) circle (2pt);		
		\draw [fill=black] (e32) circle (2pt);
		\draw [fill=black] (e41) circle (2pt);
		\draw [fill=black] (e42) circle (2pt);		

		\node[left = 1pt of e12] (text_e12) {\footnotesize{$( 1,2 )^{\epsilon}$}};
		\node[left = 1pt of e14] (text_e12) {\footnotesize{$( 1,4 )^{\epsilon}$}};
		\node[left = 1pt of e21] (text_e21) {\footnotesize{$( 2,1 )^{\epsilon}$}};
		\node[left = 1pt of e23] (text_e23) {\footnotesize{$( 2,3 )^{\epsilon}$}};
		\node[above = 1pt of e24] (text_e24) {\footnotesize{$( 2,4 )^{\epsilon}$}};
		\node[right = 1pt of e32] (text_e32) {\footnotesize{$( 3,2 )^{\epsilon}$}};
		\node[right = 1pt of e41] (text_e41) {\footnotesize{$( 4,1 )^{\epsilon}$}};
		\node[right = 1pt of e42] (text_e42) {\footnotesize{$( 4,2 )^{\epsilon}$}};

		\end{tikzpicture}
		\caption{$\mathcal{G}'$} \label{fig:pro_exp_e}
	\end{subfigure}
	\caption{An example of procedure to construct the localization interaction graph $\mathcal{G}'$ of edge agents from a given set of the subtended angle measurements.}
	\label{fig:pro_exp}
\end{figure}
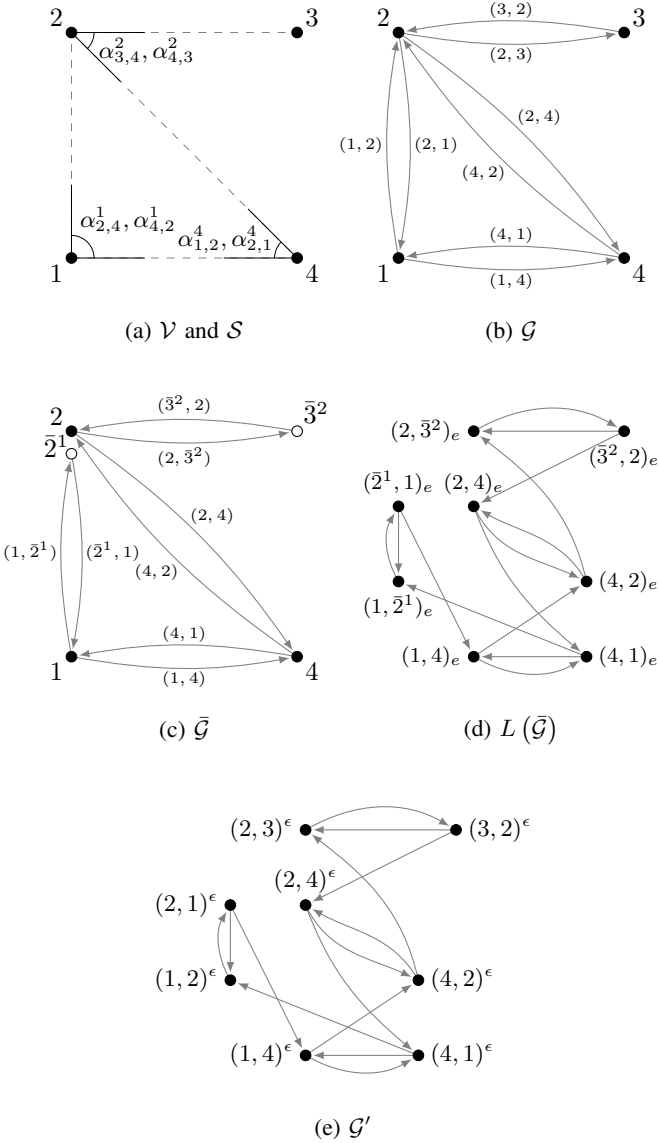

As mentioned above, the directed line graph of the edge localization graph, $L\left(\bar{\mathcal{G}}\right) = \left(  \bar{\mathcal{E}}, L\left(\bar{\mathcal{E}}\right)  \right)$, can be used as the interaction graph of edge agents.
Let $\Phi$ be an arbitrary bijective mapping from the set $\bar{\mathcal{E}}$ of edge nodes to $\mathcal{V}' = \left\{1,\ldots, \left|\bar{\mathcal{E}}\right|\right\}$, where $\left| \cdot \right|$ denotes the cardinality of a set.
Then each edge node is mapped to a distinct integer by the mapping $\Phi$.
Define the edge agent $\left( u,v \right)^{\epsilon}$ corresponding to the edge $\left( u,v \right) \in \mathcal{E}$ by $\left( u,v \right)^{\epsilon} = \Phi\left( \left( u,v \right)_{e}\right) \in \mathcal{V}'$.
Then the directed edges in $L\left(\bar{\mathcal{E}}\right)$ can be represented by their corresponding ordered pairs of the edge agents.
Let us denote a set of these ordered pairs as $\mathcal{E}'$.
Then $\mathcal{G}' = \left( \mathcal{V}', \mathcal{E}' \right)$ is a directed graph.
Finally, define the localization interaction graph of $\left| \bar{\mathcal{E}} \right|$ edge agents by the directed graph $\mathcal{G}' = \left( \mathcal{V}', \mathcal{E}' \right)$.
Example \ref{exp:procedure} simply shows the procedure to construct the localization interaction graph $\mathcal{G}'$.
\begin{example}\label{exp:procedure}
Let a graph $\mathcal{G}=(\mathcal{V}, \mathcal{E})$ with $\mathcal{V} = \left\{ 1, 2, 3, 4 \right\}$ and $\mathcal{S} = \left\{ \alpha^{1}_{2,4}, \alpha^{1}_{4,2}, \alpha^{2}_{3,4}, \alpha^{2}_{4,3}, \alpha^{4}_{1,2}, \alpha^{4}_{2,1} \right\}$, be given.
Then a procedure to construct the localization interaction graph of edge agents, $\mathcal{G}'$, from given sets $\mathcal{V}$ and $\mathcal{S}$ is illustrated in Figure \ref{fig:pro_exp}.
\end{example}
It is already assumed that an agent $u$ can obtain the information from an agent $v$ if there exists an edge $\left(u,v\right)$ in the communication graph $\mathcal{G}$.
Similarly, assume that if there is an edge $\left( k, j \right)$ in the localization interaction graph $\mathcal{G}'$, the edge agent $k$ is able to obtain the information from the edge agent $j$.
Let the edge agent $k=\left( u, v\right)^\epsilon \in \mathcal{V}'$ be assigned to the real agent $u \in \mathcal{V}$.
It is obvious that this communication between edge agents is not against the assumption of the communication graph $\mathcal{G}$.
In other words, for $k = \left( u, v\right)^{\epsilon}$ and $j = \left( v, w\right)^{\epsilon}$, the edge agent $k$ can obtain the information from the edge agent
$j$.
Furthermore, the localization interaction graph $\mathcal{G}'$ is generated in distributed manners, because each agent can create edge agents assigned to itself and interactions of the edge agents from its own communication capability in the communication graph $\mathcal{G}$ and subtended angle measurements.


\section{Edge Localization via Orientation Estimation} \label{sect:EL}
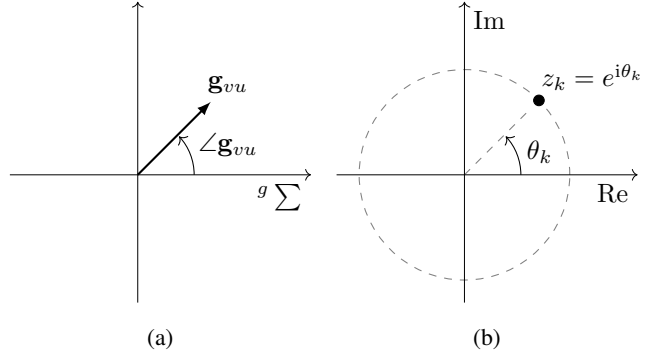
\begin{figure}[h]
	\centering
	\begin{subfigure}{0.5\columnwidth}
		\centering
		\begin{tikzpicture}
		\useasboundingbox (0,0) rectangle (4,4);
		
		\node (origin) at (1.7,1.7) {};

		\node (gvu) at ([shift=({45:1.55cm})]origin) {};

  	    \draw[-latex,black,thick] (1.7,1.7) -- (gvu);
		\centerarc[black,thin,->](origin)(0:43:0.75)

  	    \draw[->,very thin] (1.7,0) -- (1.7,4) node (yaxis) {};
		\draw[->,very thin] (0,1.7) -- (4,1.7) node (xaxis) [below left] {${}^{g}\sum$};

  		\node[black] at (2.9,2.9) {$\bold{g}_{vu}$};
  		\node[black] at (2.9,2.1) {$\angle{\bold{g}_{vu}}$};
	
		\end{tikzpicture}
		\caption{} \label{fig:orientation_e_a}
	\end{subfigure}~
	\begin{subfigure}{0.5\columnwidth}
		\centering
		\begin{tikzpicture}
		\useasboundingbox (0,0) rectangle (4,4);
		
		\node (origin) at (1.7,1.7) {};

		\node (gvu) at ([shift=({45:1.4cm})]origin) {};

		\draw[gray,thin,dashed] (origin) circle (1.4cm);
		\draw[gray,thin,dashed] (1.7,1.7) -- (gvu);

		\draw [black,fill=black] (gvu) circle (2pt);
		\centerarc[black,thin,->](origin)(0:43:0.75)

  	    \draw[->,very thin] (1.7,0) -- (1.7,4) node (yaxis) [below right] {$\rm{Im}$};
		\draw[->,very thin] (0,1.7) -- (4,1.7) node (xaxis) [below left] {$\rm{Re}$};

  		\node[black] at (3.4,3) {$z_{k} = e^{{\rm{i}}\theta_{k}}$};
  		\node[black] at (2.7,2.0) {$\theta_{k}$};
	
		\end{tikzpicture}
		\caption{} \label{fig:orientation_e_b}
	\end{subfigure}
	\caption{An example of (a) an angle of bearing vector $\bold{g}_{vu}$ and (b) an orientation angle of edge agent $k = \left(u,v\right)^\epsilon$.}
	\label{fig:orientation_e}
\end{figure}

Consider Problem \ref{prob:EL}.
Denote the edge agents' localization interaction graph by the directed graph $\mathcal{G}' = \left( \mathcal{V}', \mathcal{E}' \right)$.
Our goal is to estimate bearing vectors $\bold{g}_{vu}$ for each edge $\left(u,v\right) \in \mathcal{E}$, from given subtended angle set $\mathcal{S}$.
Let us consider an orientation, $\theta_{k}$, of an edge agent $k$ as $\angle{\bold{g}_{vu}}$ for all $k = \left(u,v\right)^\epsilon \in \mathcal{V}'$ where $\angle{\bold{g}_{vu}}\in \left[ -\pi, \pi \right)$ denotes an angle of the bearing vector $\bold{g}_{vu}$ with respect to the global coordinate frame ${}^{g}\sum$.
Then $\theta_{k}$ can be expressed by a complex number with unit magnitude and it is represented in the polar coordinate as 
$z_{k} \triangleq e^{{\rm{i}}\theta_{k}}$ for all $k \in \mathcal{V}'$.
It means that the complex variable $z_{k}$ is identical to the bearing vector $\bold{g}_{vu}$ and this relationship is illustrated in Fig. \ref{fig:orientation_e}. 
Thus the edge localization can be achieved via the orientation estimation of the edge agents.
In this section, we consider $M$ edge agents, whose interaction is represented by $\mathcal{G}'$, with a set of subtended angle measurements $\mathcal{S}$ and propose an estimation law to generate estimate $\hat{z}_k$ of $z_k$ such that $\angle\hat{z}_k \rightarrow \angle{z}_k + \angle\beta$ as $t \rightarrow \infty$ for all $k \in \mathcal{V}'$ based on measurements in $\mathcal{S}$ where $\beta$ is a common complex value.

Consider an edge agent $k=\left( u, v\right)^\epsilon$ and its neighbor $j=\left( v, w\right)^\epsilon$.
Note that the replacements of the edges by the unreachable edges guarantee disconnection of the edge agent $k$ from the edge agent $j$ in $\mathcal{G}'$, if their geometric relationship do not belong to the cases of (\ref{eq:geo_re}).
Thus, if there exists an edge agent $j \in \mathcal{N}_k$, the geometric relationship between the edge agents $k$ and $j$ should be in a case of  (\ref{eq:geo_re}) and their orientations can be represented as follows:
\begin{subequations}\label{eq:re_e_a}
\begin{numcases}{\theta_{k} = }
{\theta_{j} - \left(\alpha_{wu}^{v} + \pi\right)}	& ${\text{if } w \ne u,}$ \label{eq:re_e_a_1}\\
{\theta_{j} - \pi}									& ${\text{if } w   =  u,}$ \label{eq:re_e_a_2}
\end{numcases}
\end{subequations}
for $k=\left( u, v\right)^\epsilon$ and $j = \left( v, w\right)^\epsilon \in \mathcal{N}_k$.
Note that the edge agent $k$ can obtain and use the information of the edge agent $j$, since $k$ and $j$ are assigned to the agents $u$ and $v$, respectively, and there exists a directed edge $\left( u, v\right)$ in the communication graph $\mathcal{G}$.
We propose the following estimation law based on (\ref{eq:re_e_a}):
\begin{align}\label{eq:est_law}
{\dot{\hat z}_k}\left( t \right) = \sum\limits_{j \in \mathcal{N}_k} \left( {e^{-{\rm{i}}\theta_{jk}} {\hat z}_j\left(t\right) - {\hat z}_k\left(t\right)} \right), \; \forall k = \left(u, v\right)^\epsilon \in \mathcal{V}'		,
\end{align}
\begin{align}\label{eq:rel_ori2}
\theta_{jk} = \left.
	\begin{cases}
	{{\rm{PV}}\left(\alpha_{wu}^{v} + \pi\right)}	& {\text{if } w \ne u}	\\
	{-\pi}											& {\text{if } w = u}
	\end{cases}
\right.
\text{, for } j = \left(v, w\right)^\epsilon
\end{align}
where
\begin{align*}
{\rm{PV}}\left(\theta\right)
&\triangleq \left[\left(\theta + \pi \right) \;{\rm{mod}}\; 2\pi \right] - \pi.
\end{align*}
The estimation law (\ref{eq:est_law}) does not guarantee that $\hat{z}_k$ always stays on the unit circle even though the initial estimate $\hat{z}_k\left( t_0 \right)$ is on the unit circle.
Let the estimate $\hat{\theta}_k$ of edge agent $k$'s orientation be calculated as 
\begin{align*}
\hat{\theta}_k\left( t \right) = {\rm{PV}}\left( {\rm{arg}}\left( \hat{z}_k\left( t \right) \right) \right).
\end{align*}
Then we always can obtain $\hat{\theta}_k \left( t \right)$ although $\hat{z}_k \left( t \right)$ is not on the unit circle.
Let $\hat{\bold{z}}$ be a stacked column vector of the estimated variables defined as $\hat{\bold{z}} = \left[{ \hat{z}_{1}, \ldots, \hat{z}_{M} }\right]^T$.
Then (\ref{eq:est_law}) can be represented in the vector form as follows:
\begin{align}\label{eq:est_law2}
\dot{\hat{\bold{z}}}\left(t\right) = H \hat{\bold{z}}\left(t\right)
\end{align}
where
\begin{align*}
{\left[ H \right]}_{kj} = \left.
	\begin{cases}
	{ - \left|{\mathcal{N}_k}\right| }	& {\text{if } j = k}, \\
	{e^{-{\rm{i}}\theta_{jk}}}			& {\text{if } j \in \mathcal{N}_{k}},	\\
	{0}									& {\text{otherwise.}}
	\end{cases}
\right.
\end{align*}
The eigenvalues of $H$ can be used to prove further theorems concerning the exponential convergence of the dynamics (\ref{eq:est_law2}).
Our orientation estimation is analogous to a problem in the article \cite{lee2016distributed},
which has established that achievability of the orientation estimation problem is determined by whether the localization interaction graph has an oriented spanning tree or not.
As mentioned in Section \ref{sect:IG}, the edge localization graph $\bar{\mathcal{G}}$ is used to investigate edge localizability. 
Hence, a necessary and sufficient condition on $\bar{\mathcal{G}}$ for the localization interaction graph $\mathcal{G}'$ to have an oriented spanning tree is proved, in advance of the analysis of convergence of (\ref{eq:est_law2}).
We denote the number of oriented spanning trees of a directed graph $\mathcal{G}$ rooted at $r$ by $\kappa\left( \mathcal{G}, r \right)$.
In 1967, Knuth established a formula for the number of oriented spanning trees of a directed line graph.
The Knuth's formula is borrowed from \cite{bidkhori2011bijective}, to find the condition for that $\mathcal{G}'$ has an oriented spanning tree.
\begin{lemma}[Knuth's formula]\label{lem:knuth}
Let $\mathcal{G} = \left( \mathcal{V}, \mathcal{E} \right)$ be a directed graph in which every vertex has in-degree greater than $0$.
Then
\begin{align}\label{eq_knuth}
\kappa\left( L\left(\mathcal{G}\right), e \right) = \frac{ \kappa\left( \mathcal{G}, t\left( e \right) \right) }{d_{o}\left( h\left( e \right) \right)}
\prod\limits_{v \in \mathcal{V}}{d_{o}\left( v\right)^{d_{i}\left( v\right)-1}}, \quad\forall e \in \mathcal{E}.
\end{align}
\end{lemma}
Every vertex in the communication graph $\mathcal{G}$ has in-degree greater than $0$, since no vertices are isolated and the edge set $\mathcal{E}$ consists of pairs of two edges with opposite directions by Assumption \ref{assum:1}.
It is trivial that all vertices of the edge localization graph $\bar{\mathcal{G}}$ also  have in-degree greater than $0$.
Thus the following corollary can be derived from Lemma \ref{lem:knuth}.
\begin{corollary}\label{cor:knuth}
Let $\bar{\mathcal{G}} = \left( \bar{\mathcal{V}}, \bar{\mathcal{E}} \right)$ be the edge localization graph of the communication graph $\mathcal{G}$.
Then
\begin{align*}\label{eq_knuth2}
\kappa\left( L\left(\bar{\mathcal{G}}\right), e \right) \ge \kappa\left( \bar{\mathcal{G}}, t\left( e \right) \right), \forall e \in \bar{\mathcal{E}}.
\end{align*}
\end{corollary}
\begin{pf}
The graph $\bar{\mathcal{G}}$ is symmetric and has no vertex of in-degree less than $1$.
Thus $d_{i}\left(v\right) =  d_{o}\left(v\right) \ge 1$, for all $v \in \bar{\mathcal{V}}$.
Then (\ref{eq_knuth}) holds for the edge localization graph $\bar{\mathcal{G}}$ and the following inequality holds for all $u \in \bar{\mathcal{V}}$:
\begin{align*}
\frac{ 1 }{d_{o}\left( u \right)}\prod\limits_{v \in \bar{\mathcal{V}}}{d_{o}\left( v\right)^{d_{i}\left( v\right)-1}}
&= \frac{ {d_{o}\left( u \right)}^{d_{i}\left( u \right) - 1} }{d_{o}\left( u \right)}\prod\limits_{v \in \bar{\mathcal{V}} \setminus \left\{ u \right\}}{d_{o}\left( v\right)^{d_{i}\left( v\right)-1}}	\\
&\ge 1.
\end{align*}
Therefore,
\begin{align*}
\kappa\left( L\left(\bar{\mathcal{G}}\right), e \right) &= \frac{ \kappa\left( \bar{\mathcal{G}}, t\left( e \right) \right) }{d_{o}\left( h\left( e \right) \right)}
\prod\limits_{v \in \bar{\mathcal{V}}}{d_{o}\left( v\right)^{d_{i}\left( v\right)-1}} \\
&\ge \kappa\left( \bar{\mathcal{G}}, t\left( e \right) \right), \forall e \in \bar{\mathcal{E}}.
\end{align*}
~\hfill\text{\rule[0pt]{1.3ex}{1.3ex}}
\end{pf}
Corollary \ref{cor:knuth} is sufficient to show that if the edge localization graph $\bar{\mathcal{G}}$ has an oriented spanning tree, then its directed line graph $L\left( \bar{\mathcal{G}} \right)$  also has an oriented spanning tree.
The following lemma can be used to show its converse.
\begin{lemma}\label{cla:span_tree}
Let $\mathcal{G} = \left( \mathcal{V}, \mathcal{E} \right)$ be a directed graph.
If its directed line graph ,$L\left(\mathcal{G}\right) = \left( \mathcal{E}, L\left( \mathcal{E} \right) \right)$, has an oriented spanning tree, then $\mathcal{G}$ also has an oriented spanning tree.
\end{lemma}
\begin{pf}
Suppose that $L\left(\mathcal{G}\right)$ has an oriented spanning tree with a root $e_{r} \in \mathcal{E}$.
Let $e$ be an arbitrary edge in $\mathcal{E}\setminus \left\{ e_{r} \right\}$.
Then there exists a directed path from $e$ to $e_{r}$.
Let the directed path be $\left( e, e_1 \right), \left( e_1, e_2 \right),\dots \left( e_{n-1}, e_{n} \right), \left( e_n, e_r \right)$.
It can be represented as $\bigl( \bigl(t\bigl(e\bigr), t\bigl(e_1\bigr)\bigr), \bigl(t\bigl(e_1\bigr), t\bigl(e_2\bigr)\bigr) \bigr), \dots,$ $\bigl( \bigl(t\bigl(e_n\bigr), t\bigl(e_r\bigr)\bigr), \bigl(t\bigl(e_r\bigr), h\bigl(e_r\bigr)\bigr) \bigr)$.
Then there also exists a directed path $ \bigl(t\bigl(e\bigr), t\bigl(e_1\bigr)\bigr), \bigl(t\bigl(e_1\bigr), h\bigl(e_2\bigr)\bigr) , \dots,$ $\bigl(t\bigl(e_n\bigr), t\bigl(e_r\bigr)\bigr), \bigl(t\bigl(e_r\bigr), h\bigl(e_r\bigr)\bigr)$ in $\mathcal{G}$.
Thus, $\mathcal{G}$ has an oriented spanning tree with a root $h\left(e_r\right)$.\hfill\text{\rule[0pt]{1.3ex}{1.3ex}}
\end{pf}
Utilizing Corollary \ref{cor:knuth} and Lemma \ref{cla:span_tree}, we obtain the following result.
\begin{theorem}\label{thm:spanning_tree}
Let $\bar{\mathcal{G}} = \left( \bar{\mathcal{V}}, \bar{\mathcal{E}} \right)$ be the edge localization graph of the communication graph $\mathcal{G}$.
The edge localization graph $\bar{\mathcal{G}}$ has an oriented spanning tree if and only if
its directed line graph, $L\left(\bar{\mathcal{G}}\right) = \left( \bar{\mathcal{E}}, L\left( \bar{\mathcal{E}} \right) \right)$, has an oriented spanning tree.
\end{theorem}
\begin{pf}
$\left(\Rightarrow\right)$
Suppose that $\bar{\mathcal{G}}$ has an oriented spanning tree with a root $r$, i.e., $\kappa\left( \bar{\mathcal{G}}, r \right) \ge 1$.
Then there exists an edge $e \in \bar{\mathcal{E}}$ whose tail is $r$.
By Corollary \ref{cor:knuth}, 
\begin{align*}
\kappa\left( L\left(\bar{\mathcal{G}}\right), e \right) \ge \kappa\left( \bar{\mathcal{G}}, r \right) \ge 1.
\end{align*}
Thus $L\left(\bar{\mathcal{G}}\right)$ has an oriented spanning tree with the root $e$.

$\left(\Leftarrow\right)$
Suppose that $L\left(\bar{\mathcal{G}}\right)$ has an oriented spanning tree with a root $e_{r} \in \bar{\mathcal{E}}$.
By Lemma \ref{cla:span_tree}, $\bar{\mathcal{G}}$ has an oriented spanning tree.\hfill\text{\rule[0pt]{1.3ex}{1.3ex}}
\end{pf}
Note that the unreachable virtual vertices in the edge localization graph $\bar{\mathcal{G}}$ prevent its directed line graph $L\left(\bar{\mathcal{G}}\right)$ from having undesired connections.
The localization interaction graph $\mathcal{G}'$ is isomorphic to $L\left(\bar{\mathcal{G}}\right)$ because the vertices in $L\left(\bar{\mathcal{G}}\right)$ are merely represented as the edge agents in $\mathcal{G}'$.
Hence, $\mathcal{G}'$ is free of undesired connections between its edge agent vertices, and Corollary \ref{cor:spanning_tree} holds.
\begin{corollary}\label{cor:spanning_tree}
Let $\bar{\mathcal{G}} = \left( \bar{\mathcal{V}}, \bar{\mathcal{E}} \right)$ be the edge localization graph of the communication graph $\mathcal{G}$.
The edge localization graph $\bar{\mathcal{G}}$ has an oriented spanning tree if and only if
the localization interaction graph of edge agents, $\mathcal{G}' = \left( \mathcal{V}', \mathcal{E}' \right)$, has an oriented spanning tree.
\end{corollary}
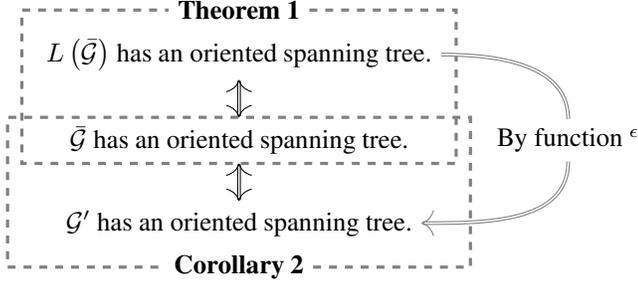
\begin{figure}[h]
	\centering

	\begin{tikzpicture}[auto]

	\node[rectangle] (L_g) {$L\left( \bar{\mathcal{G}} \right)$ has an oriented spanning tree.};
	
	\node[rectangle, below = 0.5cm of L_g] (g_bar) {$\bar{\mathcal{G}}$ has an oriented spanning tree.};

	\node[rectangle, below = 0.5cm of g_bar] (g_dot) {$\mathcal{G}'$ has an oriented spanning tree.};

	\draw[double,<->] (L_g) -- (g_bar);
	\draw[double,<->] (g_bar) -- (g_dot);

	\node[text=black, above = 0.01cm of L_g] (thm) {\textbf{Theorem \ref{thm:spanning_tree}}};
	\node[text=black, below = 0.01cm of g_dot] (cor) {\textbf{Corollary \ref{cor:spanning_tree}}};
	
	\draw[gray, dashed, very thick] (thm.east)-|([xshift=2mm]L_g.east)|-(g_bar.south)-|([xshift=-2mm]L_g.west)|-(thm.west);
	\draw[gray, dashed, very thick] (cor.east)-|([xshift=7mm]g_bar.east)|-(g_bar.north)-|([xshift=-7mm]g_bar.west)|-(cor.west);


	\node [xshift=2cm] (e) at (g_bar.east) {By function ${}^{\epsilon}$};
	\draw [gray,double,-] (L_g.east) to[out=0, in=90] (e.north);
	\draw [gray,double,->] (e.south) to[out=-90, in=0] (g_dot.east);

	\end{tikzpicture}
	\caption{Description of interrelation between Theorem \ref{thm:spanning_tree} and Corollary \ref{cor:spanning_tree}.}
	\label{fig:thm_col}
\end{figure}

Now we analyze the convergence of the dynamics (\ref{eq:est_law2}).
The following two theorems and their proofs are analogous to the ones in \cite{lee2016distributed}.
Let $\bold{z} = \bigl[{ z_1, \ldots, z_M }\bigr]^T$ be a stacked column vector.
Then the following result characterizes the eigenvalues of $H$.
\begin{theorem}\label{thm:eigen}
Zero is a simple eigenvalue of $H$ in (\ref{eq:est_law2}) with a corresponding eigenvector  $\bold{z}$ if and only if the localization interaction graph $\mathcal{G}'$ has an oriented spanning tree.
Further, every nonzero eigenvalue of $H$ has strictly negative real part.
\end{theorem}
\begin{pf}
Let $D_z = {\rm{diag}}\left({ z_1, \ldots, z_M }\right)$ be a $M \times M$ diagonal matrix.
Then $D_z$ is a invertible matrix because $z_k$, $\forall k \in \mathcal{V}'$, is on the unit circle.
By the matrix similarity, $\bar{H} = D_{z}^{-1}HD_{z}$.
Since $D_{z}$ and $D_{z}^{-1}$ are diagonal, the entries of $\bar{H}$ can be represented as follows:
\begin{align*}
\bigl[\bar{H}\bigr]_{kj} &= \bigl[{ D_{z}^{-1}HD_{z} }\bigr]_{kj} = \bigl[D_{z}^{-1}\bigr]_{k*}\bigl[{ HD_{z} }\bigr]_{*j}	\\
&= \bigl[D_{z}^{-1}\bigr]_{kk}\bigl[{ H }\bigr]_{k*}\bigl[{ D_{z} }\bigr]_{*j} = \bigl[D_{z}^{-1}\bigr]_{kk}\bigl[{ H }\bigr]_{kj}\bigl[{ D_{z} }\bigr]_{jj}
\end{align*}
where $\bigl[{ A }\bigr]_{k*}$ and $\bigl[{ A }\bigr]_{*j}$ denote $k$-th row vector and $j$-th column vector of $A$, respectively.
The diagonal entries of $\bar{H}$ is written as $\left[ \bar{H} \right]_{kk} = \bigl[D_{z}^{-1}\bigr]_{kk}\bigl[{ H }\bigr]_{kk}\bigl[{ D_{z} }\bigr]_{kk} = {z_{k}}^{-1}\bigl({ -\bigl|\mathcal{N}_{k}\bigr| }\bigr)z_{k} = -\bigl|\mathcal{N}_{k}\bigr|$, and its non-zero off-diagonal entries is written as follows:
\begin{align*}
\left[ \bar{H} \right]_{kj} = {z_{k}}^{-1} e^{-{\rm{i}}\theta_{jk}} z_{j} = e^{-{\rm{i}}\theta_{k}} e^{-{\rm{i}}\theta_{jk}} e^{{\rm{i}}\theta_{j}}.
\end{align*}
Let $k = \bigl(u,v\bigr)^{\epsilon}$ and $j = \bigl(v,w\bigr)^{\epsilon} \in \mathcal{N}_{k}$.
If $w = u$, $\left[ \bar{H} \right]_{kj} = e^{-{\rm{i}}\theta_{k}} e^{{\rm{i}}\pi} e^{{\rm{i}}\left(\theta_{k}-\pi\right)} = 1$.
Otherwise,
\begin{align*}
\left[ \bar{H} \right]_{kj} &= e^{-{\rm{i}}\theta_{k}} e^{-{\rm{i}}\theta_{jk}} e^{{\rm{i}}\theta_{j}}
= e^{-{\rm{i}}\angle{\bold{g}_{vu}}} e^{-{\rm{i}} \left({{\rm{PV}}\left(\alpha_{wu}^{v} + \pi\right)}\right) } e^{{\rm{i}}\angle{\bold{g}_{wv}}}	\\
&= e^{{\rm{i}} \left({{\rm{PV}}\left( \angle{\bold{g}_{wv}} - \angle{\bold{g}_{vu}} - \alpha_{wu}^{v} - \pi\right)} \right) } 
= e^{ {\rm{i}} \left( \pi - \pi \right) }
= 1.
\end{align*}
Thus $\bar{H}$ can be rewritten as follows:
\begin{equation*}\label{eq:bar_H}
{\left[ \bar{H} \right]}_{kj} = \left.
	\begin{cases}
	{ - \left|{\mathcal{N}_k}\right| }	& {\text{if } j = k}, \\
	{1}			& {\text{if } j \in \mathcal{N}_{k}},	\\
	{0}									& {\text{otherwise.}}.
	\end{cases}
\right.
\end{equation*}
This means that $-\bar{H}$ can be considered as a Laplacian matrix associated to a directed weighted graph which has an oriented spanning tree.
By Lemma \ref{lem:eigen}, $\bar{H}$ has a simple zero eigenvalue with right eigenvector $\bold{1}_{M} = \left[ 1, \ldots, 1\right]^{T}$ and all the other eigenvalues have strictly negative real parts.
Since $H$ and $\bar{H}$ are similar, they have same eigenvalues.
Furthermore, since $H \bold{z} = D_z \bar{H} D_z^{-1} \bold{z} = D_z \bar{H} \bold{1}_{M} = \bold{0}$, $\bold{z}$ is an eigenvector with respect to the zero eigenvalue of $H$.
\hfill\text{\rule[0pt]{1.3ex}{1.3ex}}
\end{pf}
By Theorem \ref{thm:eigen}, the equilibrium set of (\ref{eq:est_law2}) can be written as $\mathscr{E} := {\rm{span}}\left( \bold{z} \right)$.
Since the auxiliary variable $z_k$ is a complex number on the unit circle, $\hat{\bold{z}}$ should not converge to $\bold{0}$.
Therefore, define the desired equilibrium set as $\mathscr{D} := \mathscr{E} \setminus \left\{\bold{0}\right\}$.
For a matrix $A \in \mathbb{C}^{M \times M}$, let us denote the column space of $A$ by ${\rm{C}}\left( A \right)$.
Then it holds that ${\rm{C}}\left( A \right)$ is identical with the orthogonal space of null space of $A^{\ast}$, where ${}^{\ast}$ denotes the conjugate transpose.
In other words, ${\rm{C}}\left( A \right) = {\rm{null}}\left( A^{\ast} \right)^{\perp}$.
Finally, the conditions for the convergence of (\ref{eq:est_law2}) are established in the following theorem.
\begin{theorem}\label{thm:ex_con}
For the dynamics (\ref{eq:est_law2}), there exists a finite point $\bold{z}_\infty \in \mathscr{E}$
such that $\hat{\bold{z}}\left( t \right)$ exponentially converges if and only if the edge localization graph $\bar{\mathcal{G}}$ has an oriented spanning tree.
Furthermore, $\bold{z}_\infty \in \mathscr{D}$ if and only if the initial estimate $\hat{\bold{z}}\left( t_0 \right)$ is not in ${\rm{C}}\left( H \right)$.
\end{theorem}
\begin{pf}
Let $\hat{\bold{z}}\left( t \right) = D_z \bold{y}\left( t \right)$ and $\bar{H} = D_{z}^{-1}HD_{z}$, where $D_z = {\rm{diag}}\left({ z_1, \ldots, z_M }\right)$.
Then (\ref{eq:est_law2}) can be represented as 
\begin{align*}
\dot{\bold{y}}\left( t \right) &= D_{z}^{-1} \dot{\hat{\bold{z}}}\left( t \right) = D_{z}^{-1} H \hat{\bold{z}}\left(t\right) = D_{z}^{-1} H D_{z} \bold{y}\left( t \right) \\
&= \bar{H} \bold{y}\left( t \right).
\end{align*}
In Theorem \ref{thm:eigen}, it can be seen that $\bar{H}$ has zero row sum and a zero eigenvalue with corresponding right eigenvector $\bold{1}_{M} = \left[ 1, \ldots, 1\right]^{T}$.
By Lemma \ref{lem:consensus}, there exists an equilibrium set $\mathscr{E}_y \subseteq {\rm{span}}\left( \bold{1}_{M} \right) \subset \mathbb{C}^{M}$ such that $\bold{y}\left( t \right)$ exponentially converges to $\mathscr{E}_y$.
This means that there exist a finite point $\bold{y}_{\infty} \in \mathscr{E}_y$ and constants $\tau_y, \sigma_y > 0$ such that
\begin{align}\label{eq:thm3_1}
\left\| \bold{y}\left( t \right) - \bold{y}_{\infty} \right\| \le \tau_y e^{ -\sigma_y \left( t - t_0 \right)} \left\| \bold{y}\left( t_0 \right) - \bold{y}_{\infty} \right\|.
\end{align}
Let $\bold{z}_\infty = D_z \bold{y}_{\infty}$.
Since $z_k$ is on the unit circle, $D_z^{\ast} D_z = D_z D_z^{\ast} = I$.
Then, we have
\begin{align*}
\left\| D_z^{-1} \left(\hat{\bold{z}}\left( t \right) - \bold{z}_{\infty}\right) \right\|
=& \bigl\{  {\left( D_z^{-1}\left( \hat{\bold{z}}\left( t \right) - \bold{z}_{\infty} \right) \right)}^{\ast} 		\\
& \;\;\, {\left( D_z^{-1}\left( \hat{\bold{z}}\left( t \right) - \bold{z}_{\infty} \right) \right)}  \bigr\}^{\frac{{1}}{{2}}}	\\
=& \left( {\left( \hat{\bold{z}}\left( t \right) - \bold{z}_{\infty} \right)}^{\ast} {\left( \hat{\bold{z}}\left( t \right) - \bold{z}_{\infty} \right)} \right)^{\frac{{1}}{{2}}}		\\
=& \left\| \hat{\bold{z}}\left( t \right) - \bold{z}_{\infty} \right\|.
\end{align*}
From (\ref{eq:thm3_1}), the following inequality holds:
\begin{align*}
\left\| \hat{\bold{z}}\left( t \right) - \bold{z}_{\infty} \right\|
=& \left\| D_z^{-1} \left(\hat{\bold{z}}\left( t \right) - \bold{z}_{\infty}\right) \right\|
= \left\| \bold{y}\left( t \right) - \bold{y}_{\infty} \right\|		\\
\le & \tau_y e^{ -\sigma_y \left( t - t_0 \right)} \left\| \bold{y}\left( t_0 \right) - \bold{y}_{\infty} \right\|		\\
=& \tau_y e^{ -\sigma_y \left( t - t_0 \right)} \left\| D_z^{-1} \left(\hat{\bold{z}}\left( t_0 \right) - \bold{z}_{\infty}\right) \right\|		\\
=& \tau_y e^{ -\sigma_y \left( t - t_0 \right)} \left\| \hat{\bold{z}}\left( t_0 \right) - \bold{z}_{\infty} \right\|.
\end{align*}
It is trivial that $\bold{z}_{\infty} \in \mathscr{E}$, because $D_z$ is a linear map from 
$\mathscr{E}_{y}$ to $\mathscr{E}$.
It is completed to show the existence of $\bold{z}_{\infty} \in \mathscr{E}$ such that $\hat{\bold{z}}\left( t \right)$ exponentially converges.

Now we consider a solution of $\hat{\bold{z}}\left( t \right)$.
The solution can be represented as $\hat{\bold{z}}\left( t \right) = e^{H \left( t - t_0 \right)} \hat{\bold{z}}\left( t_0 \right)$.
By the similarity transformation, the Jordan form is obtained as follows:
$J = P^{-1} H P$.
Let $P$ and $P^{-1}$ be represented as $P = \left[ \bold{v}_1, \ldots, \bold{v}_M \right]$ and $P^{-1} = \left[ \bold{w}_1^{\ast}, \ldots, \bold{w}_M^{\ast} \right]^{\ast}$, respectively.
Assume that $\bold{v}_M$ is the right eigenvector corresponding to the zero eigenvalue.
Then $\bold{w}_M$ is the left eigenvector corresponding to the zero eigenvalue.
By Corollary \ref{cor:spanning_tree} and Theorem \ref{thm:eigen}, every eigenvalue of $H$ except for the zero eigenvalue has strictly negative real part.
Then the state transition matrix $e^{J \left( t - t_0 \right)}$ has the following form as $t \to \infty$:
\begin{align*}
e^{J \left( t - t_0 \right)} \to
\left[ {\begin{array}{*{20}{c}}
0&{\text{ \quad }}&{\text{ \quad }}&{\text{ \quad }}\\
{\text{ \quad }}& \ddots &{\text{ \quad }}&{\text{ \quad }}\\
{\text{ \quad }}&{\text{ \quad }}&0&{\text{ \quad }}\\
{\text{ \quad }}&{\text{ \quad }}&{\text{ \quad }}&1
\end{array}} \right]
\end{align*}
Thus a steady state solution of $\hat{\bold{z}}\left( t \right)$ is obtained as follows:
\begin{align*}
\mathop {\lim }\limits_{t \to \infty }{\hat{\bold{z}}\left( t \right)}
=& \mathop {\lim }\limits_{t \to \infty }{e^{H \left( t - t_0 \right)} \hat{\bold{z}}\left( t_0 \right)}
= \mathop {\lim }\limits_{t \to \infty }{ P e^{J \left( t - t_0 \right)} P^{-1} \hat{\bold{z}}\left( t_0 \right)}		\\
=& \bold{v}_M \bold{w}_M \hat{\bold{z}}\left( t_0 \right).
\end{align*}
Since $\bold{w}_M$ is the left eigenvector of the zero eigenvalue,
$\hat{\bold{z}}\left( t \right)$ converges to $\bold{0}$ if and only if $\hat{\bold{z}}\left( t_0 \right)$ is perpendicular to $\bold{w}_M$.
Then, this means that $\bold{w}_M \hat{\bold{z}}\left( t_0 \right) = 0$ if and only if $\hat{\bold{z}}\left( t_0 \right) \in {\rm{null}}\left(H^{\ast} \right)^{\perp}$.
Hence, $\bold{z}_\infty \in \mathscr{D}$ if and only if $\hat{\bold{z}}\left( t_0 \right) \notin {\rm{C}}\left( H \right)$.
\hfill\text{\rule[0pt]{1.3ex}{1.3ex}}
\end{pf}

By Theorem \ref{thm:ex_con}, the estimation law (\ref{eq:est_law2}) ensures that the edge localization graph has an oriented spanning tree if and only if the estimated variable $\hat{\bold{z}}$ exponentially converges to a solution with a unknown bias,
that is, each estimated value $\hat{z}_{k}$ exponentially converges to each true one ${z}_{k}$ up to a common rotation.
If we have at least a true bearing vector, $\hat{\bold{z}}$ can converge to $\bold{z}$ through (\ref{eq:est_law2}).
It is shown in Theorem \ref{thm:real_con}.
\begin{lemma}\label{cla:span_tree_w_r}
Let $\bar{\mathcal{G}} = \left( \bar{\mathcal{V}}, \bar{\mathcal{E}} \right)$ be the edge  localization graph of the communication graph $\mathcal{G}$.
If the edge localization graph $\bar{\mathcal{G}}$ has an oriented spanning tree with a root vertex $r$,
then its directed line graph, $L\left(\bar{\mathcal{G}}\right) = \left( \bar{\mathcal{E}}, L\left( \bar{\mathcal{E}} \right) \right)$, has an oriented spanning tree with a root edge $\left( r, v\right)$ for each neighbor $v$ of $r$ in $\bar{\mathcal{G}}$.
\end{lemma}
\begin{pf}
Suppose that $\bar{\mathcal{G}}$ has an oriented spanning tree with a root $r$.
Let $v_1$, $\ldots$, $v_m$ be all neighbors of $r$.
Then all edges related to $r$ in $\bar{\mathcal{G}}$ are $\left( r, v_1 \right)$, $\left( v_1, r \right)$, $\ldots$, $\left( r, v_n \right)$, and $\left( v_n, r \right)$,
since $\bar{\mathcal{G}}$ is symmetric by Assumption \ref{assum:1}.
In the directed line graph $L\left(\bar{\mathcal{G}}\right)$, there exist a directed path $\left( \left( v_1, r\right), \left( r, v_1\right) \right)$ and directed paths $\left( \left( r, v_i\right), \left( v_i, r\right) \right)$, $\left( \left( v_i, r\right), \left( r, v_1\right) \right)$ for all $1 < i \le n$.
To prove this claim, it is sufficient to show that there exists a directed path in $L\left(\bar{\mathcal{G}}\right)$ from an arbitrary edge in $\bar{\mathcal{G}}$ to $\left( v_j, r \right)$ where $1 \le j \le n$.
Let $u$ be an arbitrary vertex in $\bar{\mathcal{G}}$ except for $r$.
If $u$ is a vertex $v_j$ where $1 \le j \le n$, it is already proved.
Otherwise, there exists a directed path $\left( u, u_1 \right)$, $\ldots$, $\left( u_m, r \right)$ in $\bar{\mathcal{G}}$.
Then there also exists a directed path $\left( \left( u, u_1 \right), \left( u_1, u_2\right) \right)$, $\ldots$, $\left( \left( u_{m-1}, u_m \right), \left( u_m, r\right) \right)$ in $L\left(\bar{\mathcal{G}}\right)$.
By Definition \ref{def:line_g}, $\left( \left( u_1, u \right), \left( u, u_1\right) \right) \in L\left( \bar{\mathcal{E}}\right)$.
Thus there exists a directed path from $\left( u_1, u \right)$ to $\left( u_m, r \right)$.
This means that for arbitrary edges $\left( u, u_1 \right)$ and $\left( u_1, u \right)$, there are directed paths from them to $\left( v_j, r \right)$ where $1 \le j \le n$ in $L\left(\bar{\mathcal{G}}\right)$, respectively.
Hence the directed line graph $L\left(\bar{\mathcal{G}}\right)$ has an oriented spanning tree with a root edge $\left( r, v_1 \right)$.
This completes the proof.
\hfill\text{\rule[0pt]{1.3ex}{1.3ex}}
\end{pf}
\begin{theorem}\label{thm:real_con}
Under the dynamics (\ref{eq:est_law2}), if the edge localization graph $\bar{\mathcal{G}}$ has an oriented spanning tree with a root knowing the bearing vector from it to one of its neighbors,
the estimate $\hat{\bold{z}}$ exponentially converges to the true value $\bold{z}$.
\end{theorem}
\begin{pf}
Suppose that $\bar{\mathcal{G}}$ has an oriented spanning tree with a root $r$ knowing $\angle{\bold{g}_{{v_1}r}}$.
From Lemma \ref{cla:span_tree_w_r}, it is trivial that the localization interaction graph of edge agents, $\mathcal{G}' = \left( \mathcal{V}', \mathcal{E}' \right)$, has an oriented spanning tree with a root $l = \left( r, v_1 \right)^\epsilon$.
Since we already know $\angle{\bold{g}_{{v_1}r}}$, the edge agent $l$ does not need to update $\hat{z}_{l}$.
In other words, $\hat{z}_{l} \left( t \right) = z_{l}$, $\forall t \ge t_0$.
Thus the interaction graph of edge agents can be represented by a subgraph of $\mathcal{G}'$ by removing all edges from $l$.
Let us denote this graph by $\mathcal{G}'' = \left( \mathcal{V}', \mathcal{E}'' \right)$.
Since $\mathcal{G}'$ has an oriented spanning tree with a root $l$, $\mathcal{G}''$ also has the one.
Thus, by Theorem \ref{thm:ex_con}, the estimated variable $\hat{\bold{z}}\left( t \right)$ exponentially converges to a finite point $\bold{z}_\infty \in \mathscr{E}$.
Let $\beta \in \mathbb{C}$ such that $\bold{z}_\infty = e^{{\rm{i}}\angle{\beta}}\bold{z}$.
Then
\begin{align*}
\mathop {\lim }\limits_{t \to \infty }{\hat{\bold{z}}\left( t \right)}
=& \bold{z}_\infty
= e^{{\rm{i}}\angle{\beta}}\bold{z}.
\end{align*}
Since $\mathop {\lim }\limits_{t \to \infty }{\hat{z}_{l}\left( t \right)} = z_{l}$, we have $e^{{\rm{i}}\angle{\beta}} = 1$.
It follows that $\hat{\bold{z}}$ exponentially converges to $\bold{z}$.
\hfill\text{\rule[0pt]{1.3ex}{1.3ex}}
\end{pf}


\section{Simulation Results}\label{sec:SIM}

\begin{figure}[!h]
	\centering
	\begin{subfigure}{1\columnwidth}
		\centering
		\includegraphics[height=6cm]{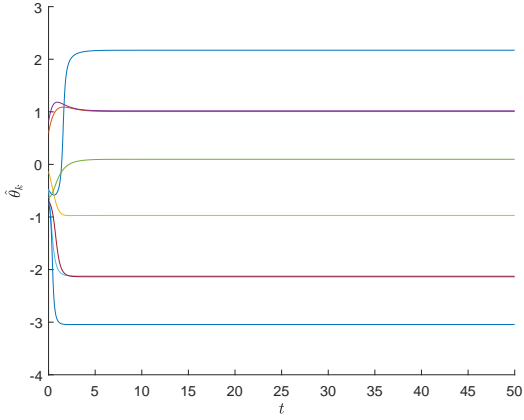}
		\caption{Convergence of orientation of each edge agent ($\hat{\theta}_k$).} \label{fig:sim_exp2_a}
	\end{subfigure}\\
	\begin{subfigure}{1\columnwidth}
		\centering
		\includegraphics[height=6cm]{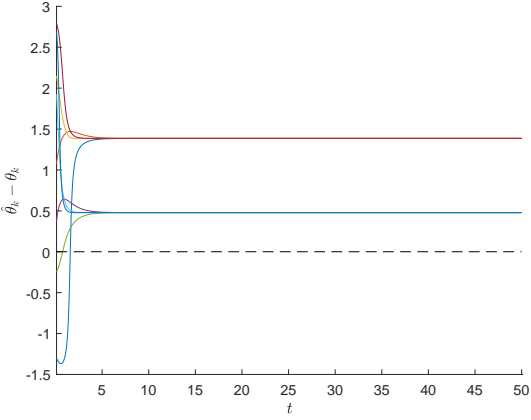}
		\caption{Consensus of orientation error of each edge agent ($\hat{\theta}_k - \theta_k$).} \label{fig:sim_exp2_a}
	\end{subfigure}
	\caption{Simulation results of Example \ref{exp:2}.}
	\label{fig:sim_exp2}
\end{figure}

\begin{figure}[!h]
	\centering
	\begin{subfigure}{1\columnwidth}
		\centering
		\includegraphics[height=6cm]{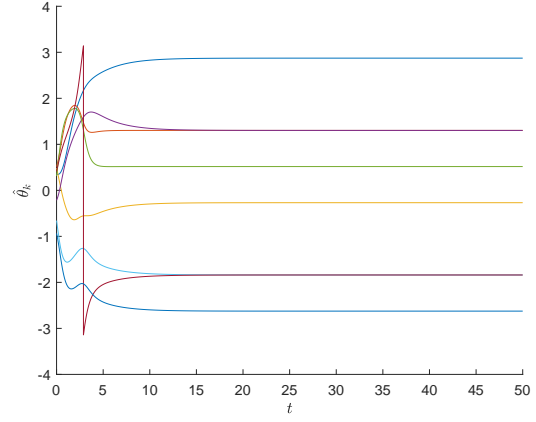}
		\caption{Convergence of orientation of each edge agent ($\hat{\theta}_k$).} \label{fig:sim_b}
	\end{subfigure}\\
	\begin{subfigure}{1\columnwidth}
		\centering
		\includegraphics[height=6cm]{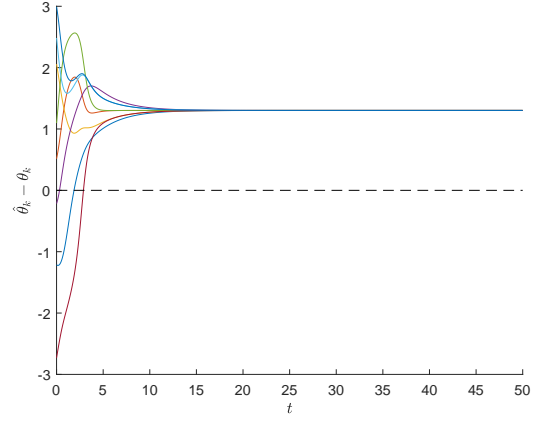}
		\caption{Consensus of orientation error of each edge agent ($\hat{\theta}_k - \theta_k$).} \label{fig:sim_c}
	\end{subfigure}
	\caption{Simulation results of Example \ref{exp:procedure}.}
	\label{fig:sim}
\end{figure}

In this section, simulation results are provided to verify the proposed edge localization method.
In the following two simulations, we consider the agents and the subtended angle measurements of Example \ref{exp:1} and Example \ref{exp:procedure}, respectively.
It can be easily seen in Figure \ref{fig:line_exp3_a} and Figure \ref{fig:pro_exp_c} that the edge localization graph $\bar{\mathcal{G}}$ of Example \ref{exp:1} does not have the oriented spanning tree, but $\bar{\mathcal{G}}$ of Example \ref{exp:procedure} does.
In the simulations, we assume that an initial orientation of each edge agent, $\hat{\bold{z}}\left( t_0 \right)$, is randomly given.
The localization interaction graphs are generated as shown in Figure \ref{fig:line_exp3_b} and Figure \ref{fig:pro_exp_d}, respectively.
For the estimation of the orientations of the edge agents, the proposed estimation law (\ref{eq:est_law}) is applied with the given $\hat{\bold{z}}\left( t_0 \right)$.
The simulation results are shown in Figure \ref{fig:sim_exp2} and Figure \ref{fig:sim}, respectively.
As illustrated in Figure \ref{fig:sim_exp2}, each orientation estimate $\hat{\theta}_k$ converges, but the orientation estimation errors do not converge to the consensus value.
Whereas, it is verified that the orientation estimation errors converge to a common bias angle $\angle\beta$ as the consensus value, in Figure \ref{fig:sim}.

In the final simulation, the simulation setting is similar to that of Figure \ref{fig:sim}, but the bearing vector $\bold{g}_{2,1}$ is assumed to be known.
It is shown that the edge localization graph $\bar{\mathcal{G}}$ has an oriented spanning tree with a root $1$ in Figure \ref{fig:pro_exp_c}.		
The edge agent $\left( 1,2 \right)^{\epsilon}$ does not need to update its orientation.
Then assume that all edges from $\left( 1,2 \right)^{\epsilon}$ are removed in the localization interaction graph $\mathcal{G}'$, as in the proof of Theorem \ref{thm:real_con}.
In this simulation, the initial orientation of $\left( 1,2 \right)^{\epsilon}$ is set as $\angle{\bold{g}}_{2,1}$ and the other ones are arbitrary given.
Figure \ref{fig:sim2} shows that the orientation estimation errors converge to zero.
\begin{figure}[!h]
	\centering
	\begin{subfigure}{1\columnwidth}
		\centering
		\includegraphics[height=6cm]{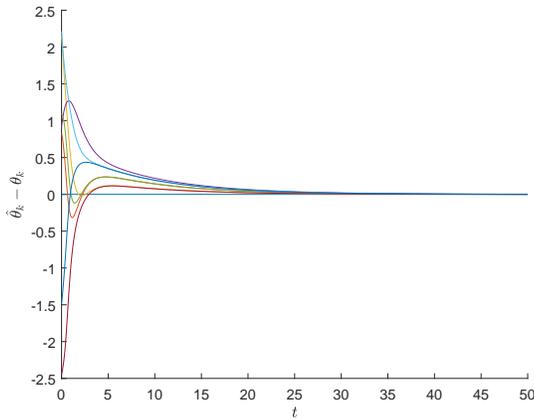}
		\caption{Consensus of orientation error of each edge agent ($\hat{\theta}_k - \theta_k$).} \label{fig:sim2}
	\end{subfigure}	
	\caption{Simulation results of Example \ref{exp:procedure} with a known bearing vector $\bold{g}_{2,1}$.} \label{fig:sim2}
\end{figure}


\section{Conclusion}\label{sec:Con}
This paper has studied the problem of estimating the bearing vectors between the agents in a multi-agent network based on only the subtended angle measurements.
We proposed two graphs; edge localization graph and localization interaction graph, to solve this problem.
A distributed estimation method via orientation estimation of virtual edge agents is proposed, utilizing these two graphs.
The proposed method ensures exponential convergence of estimation if and only if the edge localization graph has an oriented spanning tree.
Furthermore, the estimated variables exponentially converge to the true values if the edge localization graph has an oriented spanning tree with a root knowing the bearing vector from it to one of its neighbors.
As future work, the proposed method may be applied to formation control of multi-agent systems based on only the subtended angle measurements.
It is also of interest to extend the proposed method to multi-agent systems in three dimensional space.


\bibliographystyle{unsrt}
\bibliography{2018_edge_localization}

\end{document}